\documentclass[fleqn,usenatbib]{mnras}

\usepackage{newtxtext,newtxmath}
\usepackage[T1]{fontenc}

\DeclareRobustCommand{\VAN}[3]{#2}
\let\VANthebibliography\thebibliography
\def\thebibliography{\DeclareRobustCommand{\VAN}[3]{##3}\VANthebibliography}

\usepackage{graphicx}	% Including figure files
\usepackage{amsmath}	% Advanced maths commands
\usepackage{lipsum}
\usepackage{empheq}
\usepackage{mathrsfs}
\usepackage{textcomp}
\usepackage{hyperref}
\usepackage[caption=false]{subfig}
\usepackage[normalem]{ulem}
\hypersetup{colorlinks, linkcolor={blue}, citecolor={blue}, urlcolor={blue}}

\def\Eu{ \mathfrak{H} } 
\def\la{\langle}
\def\ra{\rangle}
\def\bea{\begin{eqnarray}}
\def\eea{\end{eqnarray}} 
\DeclarePairedDelimiter\abs{\lvert}{\rvert}
\newcommand{\bse}{\boldsymbol{e}}

\newcommand{\lcdm}{$\Lambda$CDM}
\newcommand{\sn}{SNe~Ia}
\newcommand{\lam}{\lambda} 

\title[Quadrupole with SNe~Ia]{The quadrupole in the local Hubble parameter: first constraints using Type Ia supernova data and forecasts for future surveys}
\author[S. Dhawan et al.]{
Suhail Dhawan,$^{1}$\thanks{E-mail: suhail.dhawan@ast.cam.ac.uk}
Antonin Borderies,$^{2}$
Hayley J. Macpherson,$^{3}$
Asta Heinesen$^{2}$
\\
$^{1}$Institute of Astronomy and Kavli Institute for Cosmology, University of Cambridge, Madingley Road, Cambridge CB3 0HA, UK\\
$^{2}$Univ Lyon, Ens de Lyon, Univ Lyon1, CNRS, Centre de Recherche Astrophysique de Lyon UMR5574, F–69007, Lyon, France\\
$^{3}$Department of Applied Mathematics and Theoretical Physics, University of Cambridge, Cambridge CB3 0WA, UK
}

\date{Accepted XXX. Received YYY; in original form ZZZ}

\pubyear{2022}

\begin{document}
\label{firstpage}
\pagerange{\pageref{firstpage}--\pageref{lastpage}}
\maketitle

\begin{abstract}
The cosmological principle  asserts that the Universe looks spatially homogeneous and isotropic on sufficiently large scales. Given the fundamental implications of the cosmological principle, it is important to empirically test its validity on various scales. In this paper, we use the Type Ia supernova (SN~Ia) magnitude-redshift relation, from both the Pantheon and JLA compilations, to constrain theoretically motivated anisotropies in the Hubble flow. 
In particular, we constrain the quadrupole moment in the effective Hubble parameter and the dipole moment in the effective deceleration parameter. 
We find no significant quadrupole term regardless of the redshift frame we use. 
Our results are consistent with the theoretical expectation of a quadrupole moment of a few percent at scales of $\sim 100$~$h^{-1}$~Mpc. We place an upper limit of a $\sim 10\%$ quadrupole amplitude relative to the monopole, $H_0$, at these scales. 
We find that we can detect a $\sim 7\%$ quadrupole moment at the 5$\sigma$ level, for a forecast low-$z$ sample of 1055 SNe~Ia. 
We find an exponentially decaying dipole moment of the deceleration parameter varies in significance depending on the redshift frame we use. In the heliocentric frame, as expected, it is detected at $\sim 3 \sigma$ significance. In the rest-frame of the cosmic microwave background (CMB), we find a marginal $\sim 2 \sigma$ dipole, however, after applying peculiar velocity corrections, 
the dipole is insignificant. 
Finally, we find the best-fit frame of rest relative to the supernovae to differ from that of the CMB.

\end{abstract}


%%%%%%%%%%%%%%%%%%%%%%%%%%%%%%%%%%%%%%%%%%%%%%%%%%

%%%%%%%%%%%%%%%%% BODY OF PAPER %%%%%%%%%%%%%%%%%%

\section{Introduction}

The {cosmological principle} %\sout{Copernican principle} 
is the backbone of modern cosmology, stipulating that the spatial distribution of matter in the Universe is homogeneous and isotropic on sufficiently large scales. A broad range of independent cosmological observations, such as fluctuations in the temperature and polarization of the cosmic microwave background \citep[CMB;][]{2020A&A...641A...6P} as well as observations of large-scale structure and matter fluctuations in the Universe --- including baryon acoustic oscillations \citep[BAO;][]{2019MNRAS.486.2184M} --- have provided compelling support for the current standard $\Lambda$ cold dark matter (\lcdm) model.
%Given that the \aht{cosmological principle} is a fundamental assumption of the standard cosmological model, it is crucial \aht{to observationally test various interpretations of the principle. }  % observational test of our best description of the large scale universe. 
Within the \lcdm\ paradigm, the interpretation of the cosmological principle is that, on large scales, distances {and light propagation} are {asymptotically} described by the {spatially} homogeneous and isotropic Friedmann-Lema\^itre-Robertson-Walker (FLRW) {general-relativistic metric solution.}  %models within general relativity. 
This is a fundamental assumption of the standard cosmological model, and it is therefore crucial to test against our observations. 

%\ahr{I propose to remove this, as the papers \citep{Goncalves2018,Goncalves2021} are not reliable in their statistical procedures. I have inserted some more classical papers in the below. } 
%\sout{Recent work has found a consistency between the transition to homogeneity and the expectations from the $\Lambda$CDM concordance model \citep{Goncalves2018,Goncalves2021}. }  
%\ahr{Erased the below, since the CMB temperature field is an integrated quantity over early+late times. }\hmr{but we can extract the primary CMB which is a probe of early times?} \ahr{Yes, you are right, however this is a highly model dependent procedure. But I think that it is fair enough to call the CMB an early Universe probe. Now reformulated in the below}
%\sout{The assumption of isotropy has been tested and validated in the early Universe.}  

%Cosmic anisotropies have been constrained with the cosmic microwave background (CMB) as an early Universe probe.
%In the early universe, the %cosmic microwave background (CMB) 
The CMB strongly  disfavours global departures from isotropy \citep[as quantified within Bianchi models; see][]{2016PhRvL.117m1302S}. Late Universe probes present complimentary constraints on the cosmological principle at small and intermediate scales, where some studies have claimed a significant detection of a dipolar anisotropy in quasar, galaxy cluster, and supernova data \citep{Secrest2021,Migkas:2021zdo,Colin:2019}. An overview of cosmic dipoles and their possible tensions with the \lcdm\ model is presented in \citet{Perivolaropoulos:2021jda}. The transition to $\lesssim 1\%$ correlations at scales $\sim $100~$h^{-1}$~Mpc has been found in Luminous Red Galaxies \citep{Hogg:2004vw}, blue galaxies \citep{2012MNRAS.425..116S}, and quasars \citep{Laurent:2016eqo} --- consistent with the \lcdm\ transition to cosmic homogeneity.
However, coherent orientations of quasar polarisation directions on 500~$h^{-1}$~Mpc scales have been detected \citep{Hutsemekers2005,Hutsemekers2014}, which could indicate %be explained by 
the existence of correlation lengths larger than expected within the \lcdm\ model.

Type Ia supernovae (\sn), owing to their standardisable luminosity, are excellent cosmological probes in the late-time Universe %whether expansion is isotropic 
\citep[see][for a review of SN~Ia cosmology]{Leibundgut2018}. The {SN~Ia} magnitude-redshift relation --- or Hubble %HM: I changed this just to Hubble diagram because we never say `Hubble-Lemaitre' diagram
%-Lemaitre 
diagram\footnote{In this paper, we focus on the relative distance measurements of SNe~Ia and do not consider the absolute luminosity calibration. Hence, we use the terms magnitude-redshift relation and Hubble %-Lemaitre 
diagram interchangeably.} --- is an independent probe of isotropy in the late Universe. A number of analyses %of the anisotropy in the magnitude-redshift relation as inferred from  
using {SN~Ia} data have found significant dipolar anisotropies in the Hubble diagram that are difficult to reconcile with \lcdm\ %, which \hmt{contradicts the FLRW model usually employed in analysis of these data}
%seem difficult to reconcile with the concordance model}  
\citep[e.g][]{2012JCAP...02..004C,Bengaly:2016,Colin:2018ghy}, while others found signals consistent with isotropy %the \lcdm\ {model}
\citep[e.g.][]{Kalus:2013,Bengaly:2015,Andrade:2018,RH2020}. 
{In these analyses,}  
%An important difference with this work is that these analyses modify 
the FLRW distance-redshift cosmography {was modified} empirically in order to allow for anisotropic signatures. % {in the }. 

In this work, we constrain anisotropic signatures in the Pantheon \citep{2018ApJ...859..101S} and Joint Lightcurve Analysis \citep[JLA;][]{2014A&A...568A..22B} SN~Ia
data using a theoretically motivated cosmographic relation. Specifically, we use the general distance-redshift cosmography from \citet{Heinesen:2020bej}, which makes no assumptions on the form of the metric tensor or field equations{. This allows for analysis of cosmological data} %and thus allows for analysis
outside of the FLRW models. We simplify this cosmography
%We reduce the number of degrees of freedom in the general cosmography using 
using the results of a recent study into local anisotropies in fully general-relativistic cosmological simulations \citep{Macpherson:2021gbh}. A key prediction of this work was that the anisotropy in the %{effective}
generalised Hubble and deceleration parameters %\aht{of the luminosity distance cosmography} 
should be dominated by a quadrupole and a dipole, respectively. 
\citet{Heinesen:2021azp} further showed that this dipole %, which dominates the deceleration parameter, 
is expected to be aligned with the local gradient in the density field. 

%Most  of 
%\hmr{All?} \ahr{Probably, at least I can't recall other investigations. } 
{Previous} studies %that have examined anisotropy using the SN~Ia magnitude-redshift relation 
have focused on constraining the dipolar signature in {SN~Ia} data. %, with varying degrees of significance in the reported values of a non-zero dipole \citep[e.g.][]{RH2020,Colin:2018ghy, 2021arXiv210812497R}.  %HM: we already cited these above
%A dipole anisotropy is indeed expected to dominate the deceleration parameter, and be 
Constraints of a quadrupole anisotropy 
%--- expected to dominate the {effective} Hubble parameter ---  
%which is expected to be present in the effective Hubble parameter of the anisotropic Hubble diagram \citep{Heinesen:2020bej,Macpherson:2021gbh} --- 
have, to the best of our knowledge, not been done. This quadrupolar anisotropy %in the effective Hubble parameter 
is of particular interest in {SN~Ia} studies since it can be constrained with relative distance measurements{, unlike the monopole, $H_0$, %of the expansion rate 
which is degenerate} %and, unlike the monopole, is %significantly less
%largely uncorrelated 
with the absolute calibration of the SN~Ia luminosity. 
%This breaking of degeneracy is 
Additionally, this quadrupolar anisotropy 
is distinct in signature from  
that of a special-relativistic boost due to our motion with respect to the CMB frame 
--- unlike a dipolar anisotropy %in the effective deceleration parameter 
which is expected to be degenerate with such a boost. %motion effects. 
%Testing for the 
The potential presence of a quadrupolar anisotropy is also interesting in light of the discrepancy in the inferred \lcdm\ Hubble parameter between early- and late-Universe probes % is beneficial since the inferred \sn\ absolute luminosity is discrepant in early and late Universe estimates 
\citep{2020A&A...641A...1P, Riess2022:h0}, since it could impact local inferences of the Hubble parameter which assume isotropy.

Recently, there have been discrepancies in the literature with respect to the significance of a dipole anisotropy in the deceleration parameter of the distance-redshift law \citep[e.g.][]{Colin:2019,RH2020}. 
With an aim to resolve this recent debate,
%in the literature, 
we also independently constrain this dipole anisotropy under various assumptions. Specifically, we study the impact of distance bias corrections, peculiar velocity (PV) corrections, and the statistical model used to define the likelihood for parameter estimation. %from the {SN~Ia} data on the inferred dipole of the deceleration parameter.  
The paper is structured as follows: in Section~\ref{sec:theory} we describe the generalised cosmographic framework and the simplifications that we make within it, in Section~\ref{sec:method} we describe the statistical methods and datasets used in our analysis. We present our results in Section~\ref{sec:results} and discuss and conclude in Section~\ref{sec:disc}.
\section{Theory}
\label{sec:theory} 
%We shall now 
In this section, we describe the theoretical basis of our cosmographic {analysis}. 
{In Section~\ref{sec:gcf}, we} review the cosmographic representation of luminosity distance %in the cosmological vicinity of an observer 
in a general space-time, and %\citep{Heinesen:2020bej}. 
%Following this, we shall consider the approximations made in the analysis of SN~Ia in the present paper. 
in Section~\ref{sec:approximations} we introduce some approximations {within this formalism, which we use in our analysis of {SN~Ia} data.
We use Greek letters to represent space-time indices which take values $0\ldots 3$, and repeated indices imply summation. We occasionally use bold-face notation and index notation interchangeably, i.e. $\boldsymbol{e}$ and $e^\mu$.

\subsection{The general cosmographic framework} 
\label{sec:gcf}
Cosmographic expressions for cosmological observables that remain agnostic about the space-time curvature --- and thus %might for instance
can incorporate arbitrary cosmic bulk flows, lensing effects, etc., in the prediction of observables --- have been examined in various works  \citep[e.g.][]{1966ApJ...143..379K,1985PhR...124..315E,1970CMaPh..19...31M,Umeh:2013UCT,Clarkson:2011br,Clarkson:2011uk,Heinesen:2020bej,Heinesen:2021qnl}.  
%{The cosmograhic frameworks as considered in \cite{Heinesen:2020bej,Heinesen:2021qnl} are particularly convenient for the analysis of data. }
Here we briefly review the general cosmographic framework for the luminosity-distance redshift relation formulated in \citet{Heinesen:2020bej}, which is particularly convenient
%appropriate 
for the analysis of {SN~Ia} data. This framework will form the basis of our anisotropic constraints.

We consider a general space-time congruence description of observers and emitters with 4--velocity field $\boldsymbol{u}$,
%\hmr{this is me being naive; but why do we sometimes use boldface in place of index notation, i.e. $\boldsymbol{u}$ vs $u^\mu$?} 
and consider observations made from a space-time event $o$. %, representing the space-time position of an observer. % of interest.} %on a single worldline of this congruence, representing the observer of interest.   
The geometric Taylor series expansion of the luminosity distance, $d_L$, to an astrophysical source {at redshift $z$ and} in direction $\boldsymbol{e}$ on the observer's sky is% given by 
\begin{equation}\label{eq:series}
    d_L(z,\boldsymbol{e}) =  d_L^{(1)}(\boldsymbol{e})\, z   + d_L^{(2)} (\boldsymbol{e})\,z^2 +  d_L^{(3)}(\boldsymbol{e}) \,z^3 + \mathcal{O}( z^4),
\end{equation} 
%where $z$ is the redshift \hmt{of the}
%to the given 
%source, and 
where the inhomogeneous and anisotropic coefficients are
\begin{equation}
\begin{aligned}
\label{eq:dLexpand2}
d_L^{(1)}(\boldsymbol{e}) &= \frac{1}{\Eu_o (\boldsymbol{e}) } \, , \qquad d_L^{(2)}(\boldsymbol{e}) =   \frac{1 - \mathfrak{Q}_o(\boldsymbol{e}) }{2 \Eu_o(\boldsymbol{e})}  \ ,\\% \qquad 
d_L^{(3)}(\boldsymbol{e}) &=  \frac{- 1 +  3 \mathfrak{Q}_o^2(\boldsymbol{e}) + \mathfrak{Q}_o(\boldsymbol{e})    -  \mathfrak{J}_o(\boldsymbol{e})   + \mathfrak{R}_o(\boldsymbol{e}) }{ 6  \Eu_o(\boldsymbol{e})}     \, ,
\end{aligned}
\end{equation} 
%are the anistropic coefficients of the cosmography, 
%as expressed in terms of the 
%\hmt{where we have defined 
and the generalised cosmological parameters are
\begin{subequations}\label{eq:paramseff}
    \begin{align}
    %\hspace*{-0.7cm} 
    \label{eq:hdef}
    \Eu(\boldsymbol{e}) &\equiv - \frac{1}{E^2}     \frac{ {\rm d} E }{{\rm d} \lambda}   \, ,\\ % \qquad 
    \mathfrak{Q}(\boldsymbol{e})  &\equiv - 1 - \frac{1}{E} \frac{     \frac{ {\rm d} \Eu}{{\rm d} \lambda}    }{\Eu^2}   \, , \\ %\qquad
    \mathfrak{R}(\boldsymbol{e}) &\equiv  1 +  \mathfrak{Q}  - \frac{1}{2 E^2} \frac{k^{\mu}k^\nu R_{\mu \nu} }{\Eu^2}   \, , \\% \qquad
    \mathfrak{J}(\boldsymbol{e})  &\equiv   \frac{1}{E^2} \frac{      \frac{  {\rm d^2} \Eu}{{\rm d} \lambda^2}    }{\Eu^3}  - 4  \mathfrak{Q}  - 3 \, .
    \end{align}
\end{subequations} 
Here, $E = -u^\mu k_\mu$ is the observed photon energy, $\lambda$ is the affine parameter of the geodesic, $\frac{ {\rm d}  }{{\rm d} \lambda} \equiv k^\mu \nabla_\mu$ is the directional derivative along the incoming null ray, 
$R_{\mu \nu}$ is the Ricci curvature of the space-time, and %the operator 
the photon 4--momentum %$\boldsymbol{k}$ 
can be decomposed as $k^\mu = E(u^\mu - e^\mu)$. 
{The inverse energy function, $1/E$, replaces the FLRW scale factor in the luminosity distance cosmography for a general space-time, and can thus be thought of as a natural ``scale-factor'' %generalisation 
on the observer's past light cone. }
{The parameters $\{\Eu,\mathfrak{Q},\mathfrak{J},\mathfrak{R}\}$ 
%can be thought of as 
represent inhomogeneous, anisotropic generalisations of the FLRW Hubble, deceleration, jerk, and curvature parameters. }
We shall {therefore} refer to $\{\Eu,\mathfrak{Q},\mathfrak{J},\mathfrak{R}\}$ as the \emph{effective} observational Hubble, deceleration, jerk and curvature parameters. These effective cosmological parameters include information about regional kinematics and curvature effects; for instance bulk flow motions or the lensing of photons. 
In the strictly homogeneous and isotropic limit of \eqref{eq:paramseff}, % the equations, 
{we recover} the well-known FLRW cosmographic results of \cite{Visser:2003vq}. % are recovered. 

The anisotropic signatures of the effective cosmological parameters % might 
can be %conveniently 
represented by multipole series in the direction vector $\boldsymbol{e}$. For instance, the effective observational Hubble parameter can be expanded 
%in $\boldsymbol{e}$ 
as follows\footnote{See \cite{Heinesen:2020bej} for details on regularity requirements of the series.}
\begin{equation}
\label{def:Eevolution}
    \Eu(\boldsymbol{e}) = \frac{1}{3}\theta  - e^\mu a_\mu + e^\mu e^\nu \sigma_{\mu \nu}   \,  , 
\end{equation} 
where $\theta$ is the volume expansion rate of the observer congruence, $\sigma_{\mu \nu}$ is its volume-preserving deformation (shear), and 
%$\boldsymbol{a}$ 
$a^\mu$ is its 4--acceleration. % of the observer congruence. 
%We emphasise that the representation \eqref{def:Eevolution} is \emph{exact}, and that the truncation in the multipole expansion at the quadrupolar level is a general feature of any congruence description. 
We emphasise that the {multipole} expansion \eqref{def:Eevolution} is \emph{exact}, and represents all contributions of anisotropy to the effective Hubble parameter. 
%In a similar fashion, 
{The effective deceleration parameter can be decomposed into multipoles in a similar way, and reads}  
%as a truncated multipole series in $\boldsymbol{e}$ in the following way: 
\begin{equation}
\begin{aligned}
\label{q}
\mathfrak{Q}(\boldsymbol{e} ) = - 1& -  \frac{1}{\Eu^2(\boldsymbol{e} )} \bigg(
\overset{0}{\mathfrak{q}}   +  e^\mu  \overset{1}{\mathfrak{q}}_\mu   +    e^\mu e^\nu   \overset{2}{\mathfrak{q}}_{\mu \nu}     \\
&+    e^\mu e^\nu e^\rho \overset{3}{\mathfrak{q}}_{\mu \nu \rho}    +   e^\mu e^\nu e^\rho e^\kappa  \overset{4}{\mathfrak{q}}_{\mu \nu \rho \kappa} \bigg)     \, , 
\end{aligned}
\end{equation}
with coefficients 
\bea
\label{qpoles}
&& \overset{0}{\mathfrak{q}} \equiv  \frac{1}{3}   \frac{ {\rm d}  \theta}{{\rm d} \tau} + \frac{1}{3} D_{   \mu} a^{\mu  } - \frac{2}{3}a^{\mu} a_{\mu}    - \frac{2}{5} \sigma_{\mu \nu} \sigma^{\mu \nu}    \, , \nonumber \\ 
&& \overset{1}{\mathfrak{q}}_\mu \equiv  - \frac{1}{3} D_{\mu} \theta   -  \frac{2}{5}   D_{  \nu} \sigma^{\nu }_{\;  \mu  }   -  \frac{ {\rm d}  a_\mu }{{\rm d} \tau}  + a^\nu \omega_{\mu \nu}  +  \frac{9}{5}  a^\nu \sigma_{\mu \nu}     \, , \nonumber \\
&& \overset{2}{\mathfrak{q}}_{\mu \nu}  \equiv     \frac{ {\rm d}  \sigma_{\mu \nu}   }{{\rm d} \tau} +   D_{  \la \mu} a_{\nu \ra } + a_{\la \mu}a_{\nu \ra }     - 2 \sigma_{\alpha (  \mu} \omega^\alpha_{\; \nu )}   - \frac{6}{7} \sigma_{\alpha \la \mu} \sigma^\alpha_{\; \nu \ra }   \, , \nonumber \\ 
 && \overset{3}{\mathfrak{q}}_{\mu \nu \rho}  \equiv -  D_{ \la \mu} \sigma_{\nu   \rho \ra }    -  3  a_{ \la \mu} \sigma_{\nu \rho \ra }    \, , \nonumber\\% \qquad \quad  
 && \overset{4}{\mathfrak{q}}_{\mu \nu \rho \kappa}  \equiv  2   \sigma_{\la \mu \nu } \sigma_{\rho \kappa \ra} \, , 
\eea 
where $\frac{ {\rm d} }{{\rm d} \tau} \equiv u^\mu \nabla_\mu$ is the directional derivative along the observer 4--velocity field and $\omega_{\mu \nu}$ is the vorticity tensor describing the rotation of the observer congruence. 
Triangular brackets $\la \ra$ around indices single out the traceless and symmetric part of the tensor in those indices. 
In this work, we focus on the effective Hubble and deceleration parameters, and we therefore refer the reader to \cite{Heinesen:2020bej} for the {multipole series expressions} for $\mathfrak{J}$ and $\mathfrak{R}$. 

%The formalism presented in Section~\ref{sec:gcf} 
{This formalism} has the advantage of being general, and can in principle be applied for a fully model-independent data analysis {of standardisable candles}. However, as detailed in \cite{Heinesen:2020bej}, such an analysis 
would require the determination of 61 independent degrees of freedom. This level of constraining power is %in practice 
not achievable with current SN~Ia catalogues, and assumptions are therefore necessary to apply the framework to available data. 
{In the next section, we will make {physically motivated} approximations to simplify the above multipole expansions for our analysis.}

\subsection{Approximations} \label{sec:approximations}

We consider %the case of a 
geodesic astrophysical sources, such that %congruence description, such that the fluid acceleration 
$\boldsymbol{a} = \boldsymbol{0}$, and consider
%We furthermore {consider 
scales where expansion dominates over anisotropic deformation of space, such that % {constrain} %consider the case where 
%the anisotropic 
shear and vorticity are %to be %deformation of the space-time congruence are 
subdominant to the isotropic expansion. More specifically, we assume %such that: 
$\abs{e^\mu e^\nu \sigma_{\mu \nu}}_o \ll \theta_o$, $\abs{e^\mu e^\nu \sigma_{\alpha \mu} \sigma^{\alpha}_{\, \nu}}_o \ll \theta^2_o$, $\abs{e^\mu e^\nu \sigma_{\alpha \mu} \omega^{\alpha}_{\, \nu}}_o \ll \theta^2_o$, and $\abs{ e^\mu e^\nu  {\rm d}  \sigma_{\mu \nu} /  {\rm d} \tau}_o \ll \theta^2_o$ for all directions  %$\boldsymbol{e}_o$ 
on the observer's sky. 
However, we shall \emph{not} impose any smallness conditions on the spatial gradients of the kinematic variables. In particular $\abs{e^\mu D_{\mu} \theta}_o$ and $\abs{e^\mu e^\nu e^\sigma D_{\mu} \sigma_{\nu \sigma}}_o$ might be of order $\theta^2_o$ or larger. Indeed, for weak field expansions in cosmology, spatial gradients tend to increase the order of magnitude of the metric perturbation on scales below the Hubble horizon \citep{Rasanen:2009mg,Rasanen:2010wz,Buchert:2009wj}. 

Under the above weak-anisotropy approximations, including only the leading order anisotropic terms in \eqref{def:Eevolution} and \eqref{q} leads to
\begin{equation}
\label{def:EevolutionApp}
    H(\boldsymbol{e}) = \frac{1}{3}\theta + e^\mu e^\nu  \sigma_{\mu \nu}, 
\end{equation} 
and 
\begin{equation} 
\label{qApp}
    q(\boldsymbol{e}) = - 1 -  \frac{\overset{0}{{q}}   +  e^\mu  \overset{1}{{q}}_\mu   +    e^\mu e^\nu e^\rho \overset{3}{{q}}_{\mu \nu \rho} }{\frac{1}{9} \theta^2}, 
\end{equation}
with coefficients 
\newcommand{\be}{\boldsymbol{e}}
\begin{equation} 
\label{qpolesApp}
\overset{0}{{q}} =  \frac{1}{3}   \frac{ {\rm d}  \theta}{{\rm d} \tau}     \, , \quad \overset{1}{{q}}_\mu =  - \frac{1}{3} D_{\mu} \theta   -  \frac{2}{5}   D_{  \nu} \sigma^{\nu }_{\;  \mu  }      \, , \quad  \overset{3}{{q}}_{\mu \nu \rho}  = -  D_{ \la \mu} \sigma_{\nu   \rho \ra },
\end{equation}
{where we have defined $\Eu(\be) \rightarrow H(\be)$ and $\mathfrak{Q}(\be) \rightarrow q(\be)$ in this limit of weak anisotropy.}
In the following analysis, we shall further assume that the traceless part of $e^\mu e^\nu e^\sigma D_{\mu} \sigma_{\nu \sigma}$ (incorporated in $\overset{3}{\mathfrak{q}}_{\mu \nu \rho}$) is subdominant to its trace (incorporated in $\overset{1}{\mathfrak{q}}_\mu$), and thus set $\overset{3}{\mathfrak{q}}_{\mu \nu \rho} = 0$. 
We shall make this assumption from a practical viewpoint because of the sparsity of the {data we use (see Section~\ref{sec:data}),} % \aht{investigated SN~Ia catalogues}, 
making it unrealistic to resolve an octupole feature on the sky. 
For the same reason, we shall also not investigate anisotropic terms in the higher-order effective observational parameters $\mathfrak{J}$ and $\mathfrak{R}$. 
For future surveys with {more data and improved sky coverage, 
we will be able to} include a more complete hierarchy of anisotropies. 

We note that $D_{  \nu} \sigma^{\nu }_{\;  \mu  }  = \frac{2}{3}D_{\mu} \theta $ for a general-relativistic irrotational dust space-time \citep{Buchert:1999er}, which in this case makes the interpretation of the dipole term, $\overset{1}{{q}}_\mu$, in \eqref{qpolesApp} %in the deceleration parameter 
clearly related to %$\overset{1}{\mathfrak{q}}_\mu$ %clear as 
the spatial gradient of the expansion rate, $\theta$. Furthermore, spatial gradients of the expansion rate are expected to be proportional to spatial gradients of the density field in large-scale cosmological modelling \citep{Heinesen:2021azp}, which implies that we expect the dipole in the effective deceleration parameter to be aligned with the spatial gradient of the local density field.

\subsection{Anisotropic cosmography} 
\label{sec:cosmography} 

The JLA catalogue covers a wide range of redshifts $0.01 \lesssim z \lesssim 1.3$. 

As discussed in Appendix~A of \citet{Macpherson:2021gbh}, cosmography for anisotropic space-time models is best suited for narrow redshift intervals.
Thus, in order to apply the above formalism to a wide redshift range, we shall allow for decaying anisotropic {signatures} with redshift. This results in a cosmography which might be highly anisotropic at small scales, but which transitions into the well-known isotropic cosmography at the largest scales of observation. 

With the simplifications given in the previous section, the cosmographic expansion of $d_L$ becomes
\begin{align}\label{eq:dL_q0_dipole}
%D_L(z) = \frac{c z}{H} \{1 + \frac{(1-q)z}{2}  + \frac{-(1 - q - 3q^2 + j_0 - \Omega_{\rm K})}{6} z^2 \}
    d_L(z,\bse) &= \frac{z}{H(\bse)} \bigg\{1 + \frac{[1-q(\bse)]z}{2} \\
    &+ \frac{-[1 - q(\bse) - 3q(\bse)^2 + j_0 - \Omega_{\rm K}]}{6} z^2 \bigg\}, \notag
\end{align}
where 
we have applied the notation $\mathfrak{R}(\be)\rightarrow \Omega_{k}$ and $\mathfrak{J}(\be)\rightarrow j_0$ from FLRW cosmography, since we are only considering the monopolar contributions to $\mathfrak{R}(\be)$ and $\mathfrak{J}(\be)$ in this analysis.
Since $j_0$ and $\Omega_{\rm K}$ are degenerate in the expression \eqref{eq:dL_q0_dipole}, we will constrain the combination $j_0 - \Omega_{\rm K}$. 
{We express the anisotropic deceleration parameter by re-writing \eqref{qApp} as}
\begin{equation}
    %q = q_m + {\bf q_d} \cdot \hat{n} \mathcal{F}(z, S)
    q(\bse) = q_m + {\bf q_d} \cdot \bse \, \mathcal{F}_{\rm dip}(z, S),
    \label{eq:q_z_dipole}
\end{equation}
where $q_m$ and ${\bf q_d}$ are the monopole and dipole components,
respectively, 
and $\mathcal{F}_{\rm dip}(z, S)$ describes the scale {dependence of the dipole}. The ansatz \eqref{eq:q_z_dipole} for the deceleration parameter coincides with that of \footnote{In the analysis of \cite{Colin:2018ghy}, the direction of the source is indicated by the variable $\hat n$, which in our notation reads $\bse$.} \cite{Colin:2018ghy}. 

We now express the anisotropic Hubble parameter {by re-writing \eqref{def:EevolutionApp} as }
\begin{align}
    %H = (H_m)\cdot \left(1 + \left(\lambda_1 \cdot {\rm cos}^2\theta_1+ \lambda_2 \cdot {\rm cos}^2\theta_2 - (\lambda_1 + \lambda_2) \cdot {\rm cos}^2\theta_3\right)\mathcal{F} \right)
    %H(\bse) &= H_m \bigg\{1 + \bigg[\lambda_1 \cdot {\rm cos}^2\theta_1+ \lambda_2 \cdot {\rm cos}^2\theta_2 \\
    %&- (\lambda_1 + \lambda_2) \cdot {\rm cos}^2\theta_3\bigg]\mathcal{F}_{\rm quad}(z,S) \bigg\}
    H(\bse) &= H_m + {\bf H_q} \cdot \bse \bse \,  \mathcal{F}_{\rm quad}(z,S)
\label{eq:h_z_quadrupole}
\end{align}
where $H_m = H_0$ and ${\bf H_q}$ are the the monopole and quadrupole components, respectively,  % of $H(\bse)$, 
and $\mathcal{F}_{\rm quad}(z,S)$ describes the scale dependence of the quadrupole. We denote {the} eigenvalues of the {normalised quadrupole tensor ${\bf H_q}/H_0$} as $\lambda_1$, $\lambda_2${, and $\lambda_3 = - \lambda_1 - \lambda_2$}, and the eigendirections as {$\boldsymbol{\theta}_1$, $\boldsymbol{\theta}_2$, and $\boldsymbol{\theta}_3$}, such that
\begin{equation}
\begin{aligned}
    H(\bse) &= H_m \bigg\{1 + \bigg[\lambda_1 \cdot {\rm cos}^2\theta_1+ \lambda_2 \cdot {\rm cos}^2\theta_2 \\
    &- (\lambda_1 + \lambda_2) \cdot {\rm cos}^2\theta_3\bigg]\mathcal{F}_{\rm quad}(z,S) \bigg\},
\end{aligned}
\end{equation}
where $\theta_i$ are the angular separations between the coordinates of the supernova and the eigendirections $\boldsymbol{\theta}_i$. 
%From our constraints, 
In the following analysis, we will quote the total \textit{amplitude} of the quadrupole component of $H$ (relative to the monopole $H_m$) as the norm of the tensor ${\bf H_q}$ multiplied by the decay function $\mathcal{F}$, namely 
\begin{align}
    A_q &= || {\bf H_q} || \,\mathcal{F}_{\rm quad}(z,S) \\
    & = \sqrt{\lambda_1^2 + \lambda_2^2 + \left(\lambda_1 + \lambda_2\right)^2}\, \mathcal{F}_{\rm quad}(z,S), \label{eq:ampcalc}
\end{align}
for some redshift scale $z$. 

{Previous} anisotropy searches in the literature {have employed various} forms of $\mathcal{F}$, including constant, linear{, and exponential laws} in redshift { \citep{Colin:2018ghy}}. Recent Bayesian model comparison studies strongly disfavour a  
 constant-in-redshift dipole in data over scales of $\sim 1$Gpc %{were achieved in} %essentially ruled it out 
\citep{2021arXiv210812497R}.
The redshift range of the survey is important for the interpretation of the (amplitude of) anisotropic coefficients in the cosmographic fit \citep{Macpherson:2021gbh}. The datasets that we investigate span redshifts up to $z \sim 1$, and we thus expect a transition towards an approximately isotropic cosmography for the most distant \sn\ in the sample.
We therefore assume {the} exponential form $\mathcal{F}(z, S) = {\rm exp} (\frac{-z}{S})$, where $S$ is the decay scale, for both the dipole in $q$ and the quadrupole in $H$. For our fiducial case, we fit the scales for the dipole and quadrupole as distinct parameters $S_d$ and $S_q$, respectively. We also fit two other parametrisations of $\mathcal{F}$ in the quadrupole: a step function with a fixed width in redshift and the exponential model with a fixed decay scale $S_q$. 
The former is expressed as $\mathcal{F}(z, z_{\rm step})$, where $\mathcal{F}(z\leq z_{\rm step}, z_{\rm step} )=1$ and $\mathcal{F}(z>z_{\rm step}, z_{\rm step})=0$.

\section{Methodology and data}
\label{sec:method} 

The distance modulus of an astrophysical object is defined in terms of the absolute magnitude, $M$, of the object and the apparent magnitude, $m$, as measured by the observer. SNe~Ia corrected magnitudes are inferred in the $B$-band and are related to the luminosity distance, $d_L$, in the following way  
%Theoretically, the distance modulus predicted by the homogeneous and isotropic, flat Friedman-Robertson-Walker (FRW) universe is given by
\begin{equation}
    \mu \equiv m_B^{*} - M_B = 5\, \mathrm{log_{10}} \left( \frac{d_L}{10\, \mathrm{Mpc}} \right) + 25 \, . 
\label{eq:mu_sne}
\end{equation} 
%In practice, detected magnitudes of objects are not bolometric and supernovae are not perfect standard candles, but are associated with variations in their intrinsic properties. 
Observationally, the standardized
\sn\ peak magnitude $m^*_B$ is estimated from correcting the peak apparent magnitude, $m_B$, for correlations with the light curve width, $x_1$, and colour, $c$, to infer the distance modulus using the following relation \citep{1998A&A...331..815T,2014A&A...568A..22B} 
\begin{equation}
    \mu_{obs} = m_B - (M_B - \alpha x_1 + \beta c),
\label{eq:obs_distmod}
\end{equation}
where $M_B$ is the {mean} absolute magnitude of the \sn\ in the B-band.\footnote{These corrections are already applied to the Pantheon dataset, however, we test their impact on the cosmological parameters in Appendix~\ref{appx:pantheon_corr} and find no correlation.} Following \cite{2014A&A...568A..22B}, we apply a step correction, $\Delta_\mathrm{M}$, depending on the host galaxy stellar mass. This step correction accounts for the observation that \emph{after} stretch and colour correction, the SNe~Ia in high mass hosts are on average brighter than those in low mass hosts \citep[e.g., see,][]{2014A&A...568A..22B}. We note that $\alpha$, $\beta$, $M_B$ and $\Delta_\mathrm{M}$ are nuisance parameters in the fit for the cosmology. \footnote{We note that the $\Delta_\mathrm{M}$ parameter is only implemented in the $
\chi^2$ method discussed below.}
%\ahr{In Antonin's analysis we dont have $\Delta_\mathrm{M}$ as a nuisance parameter of the analysis, clarify this. Also explain how $\Delta_\mathrm{M}$ enters in the analysis in footnote.  }
We insert the geometrical prediction for $d_L$ given in \eqref{eq:dL_q0_dipole} into \eqref{eq:mu_sne} for our anisotropic analysis. We emphasize that the parameters describing the anisotropies that we constrain, e.g. $H_q$ and $q_d$, are not degenerate with the SN~Ia absolute $B$-band magnitude. The monopole of the Hubble parameter, $H_m$, \emph{is} however degenerate with $M_B$ and is thus not constrained by our analysis. %unlike the case for the monopole of the Hubble parameter.

%of data under the approximations outlined in Section~\ref{sec:theory}. 
%We note that 
%Any non-zero value of the quadrupole ${\bf H_q}$ %in \eqref{eq:h_z_quadrupole} 
%introduces a breaking of the degeneracy of the mean {B-band} absolute magnitude of the \sn\ and the effective Hubble parameter $H(\bse)$, which is normally present in isotropic analysis.

In order to ensure that our results are robust, we constrain the anisotropic parameters using two independent statistical methods, namely a constrain{ed} $\chi^2$ method (detailed in Section~\ref{sec:chisq}) and a maximum {likelihood} estimation method (detailed in Section~\ref{sec:MLE}). 
%which we outline in the following sections.  
%\ahr{I erased the below remark. I think that the above suffices for the introduction to the analysis. Our further procedure will be clear in the results section.   } 
%For both methods, we constrain the dipole in the deceleration parameter and the quadrupole in the Hubble parameter separately.
%Specifically, we set the dipolar contribution in $q(\be)$ to zero in order to constrain the quadrupole in the Hubble parameter. Similarly, we set the quadrupolar contribution in $H(\be)$ to zero in order to constrain the dipole in the deceleration parameter.

\subsection{Constrained $\chi^2$ method}\label{sec:chisq}

{We use the} observed distance modulus \eqref{eq:obs_distmod} %might be used 
to constrain a parametrised geometric prediction of the distance modulus by constructing the test statistic with an assumed $\chi^2$-distribution, namely  
\begin{equation}
    \chi_{\mathrm{SN}}^2 = \boldsymbol{\Delta}^T \boldsymbol{C}_{\mathrm{SN}}^{-1} \boldsymbol{\Delta}, \label{eq:chisq}
\end{equation} 
where $\boldsymbol{\Delta} = \boldsymbol{\mu}_{\text{th}} - \boldsymbol{\mu}_{\text{obs}}$ is the residual vector of the 
%theoretically prescribed 
theoretical distance moduli $\boldsymbol{\mu}_{\text{th}}$ and observed distance moduli $\boldsymbol{\mu}_{\text{obs}}$ of the sample, and {$\boldsymbol{C}_{\mathrm{SN}}$} is the covariance matrix of the observations. 
%In the present analysis of the JLA sample, 
%{In our analysis,} 
We use \eqref{eq:dL_q0_dipole} and \eqref{eq:mu_sne}, %%to compute the geometric distance modulus vector 
in place of the FLRW relation usually employed in isotropic analyses, to compute $\boldsymbol{\mu}_{\text{th}}$. The estimate of the complete covariance matrix, {$\boldsymbol{C}_{\mathrm{SN}}$}, {is} described in \cite{2014A&A...568A..22B}.  We use \texttt{PyMultiNest} \citep{2014A&A...564A.125B}, a python wrapper to \texttt{MultiNest} \citep{2009MNRAS.398.1601F}, to derive the posterior distribution of the anisotropic parameters.

\subsection{Maximum Likelihood Estimation (MLE)}\label{sec:MLE}

We use the likelihood construction of \cite{Nielsen:2015pga} \citep[see also Section~3.1 of][]{Dam:2017xqs}, in which the \sn\ are assumed to be standardisable such that the intrinsic magnitude, colour, and shape parameters describing the lightcurve of the individual \sn\
%are assumed to 
may be drawn from identical Gaussian distributions. 
In the final likelihood, it is thus the expectation value of the intrinsic Gaussian distributions that enter in the relation \eqref{eq:obs_distmod}, and \emph{not} the measured {SN~Ia} parameters themselves, which are subject to scatter. 
As in Section~\ref{sec:chisq}, %the constrained $\chi^2$ method, 
the geometric prediction of the distance modulus $\boldsymbol{\mu}$ is given by the cosmography in Section~\ref{sec:theory}, and the experimental covariance matrix of the likelihood function is that described in \cite{2014A&A...568A..22B}. 
In addition to the cosmographic parameters of interest, the analysis contains a number of nuisance parameters. Namely, the coefficients $\alpha$ and $\beta$ of the relation \eqref{eq:obs_distmod}, and the parameters describing the hypothesised Gaussian distributions of the true {SN~Ia} lightcurve parameters. 
In the likelihood optimisation we will marginalise over these nuisance parameters. 
}

\begin{figure}
    \centering
    \includegraphics[width=0.5\textwidth]{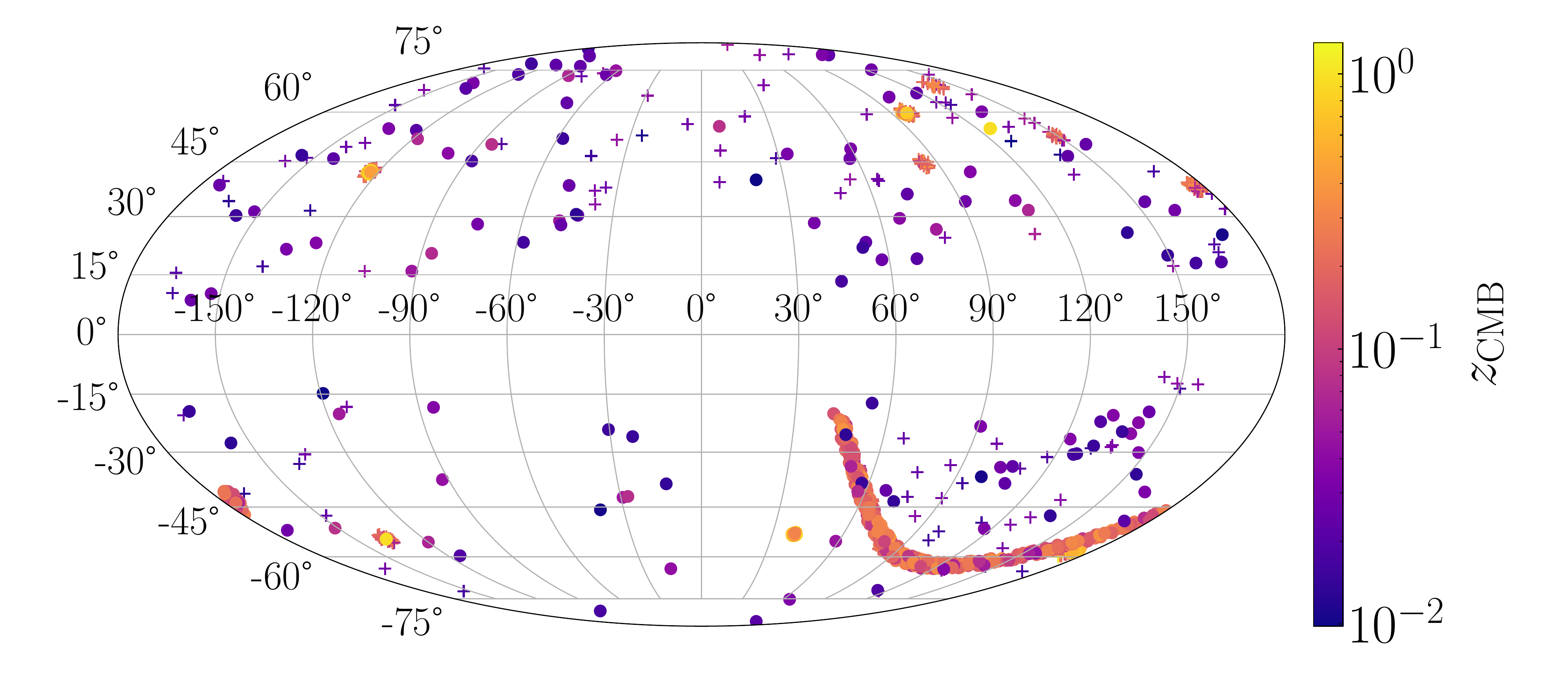}
    \caption{Sky coverage of the supernova samples used in this work. Crosses show the sky location of Pantheon supernovae and circles show the JLA supernovae in galactic coordinates $(l,b)$. Each point is coloured according to the redshift of that supernova in the CMB frame, $z_{\rm CMB}$.}
    \label{fig:SNe_direc}
\end{figure}

\begin{figure*}
    \centering
    \includegraphics[width=.48\textwidth, trim = 0 10 0 10]{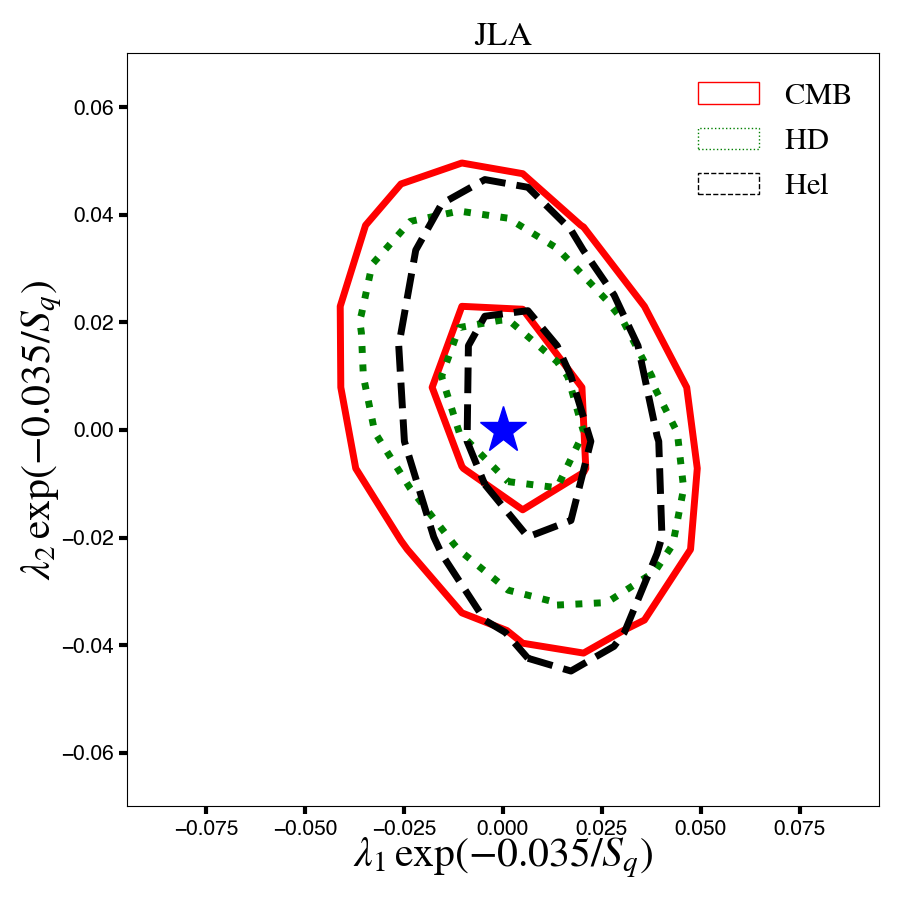}
    \includegraphics[width=.48\textwidth, trim = 0 10 0 10]{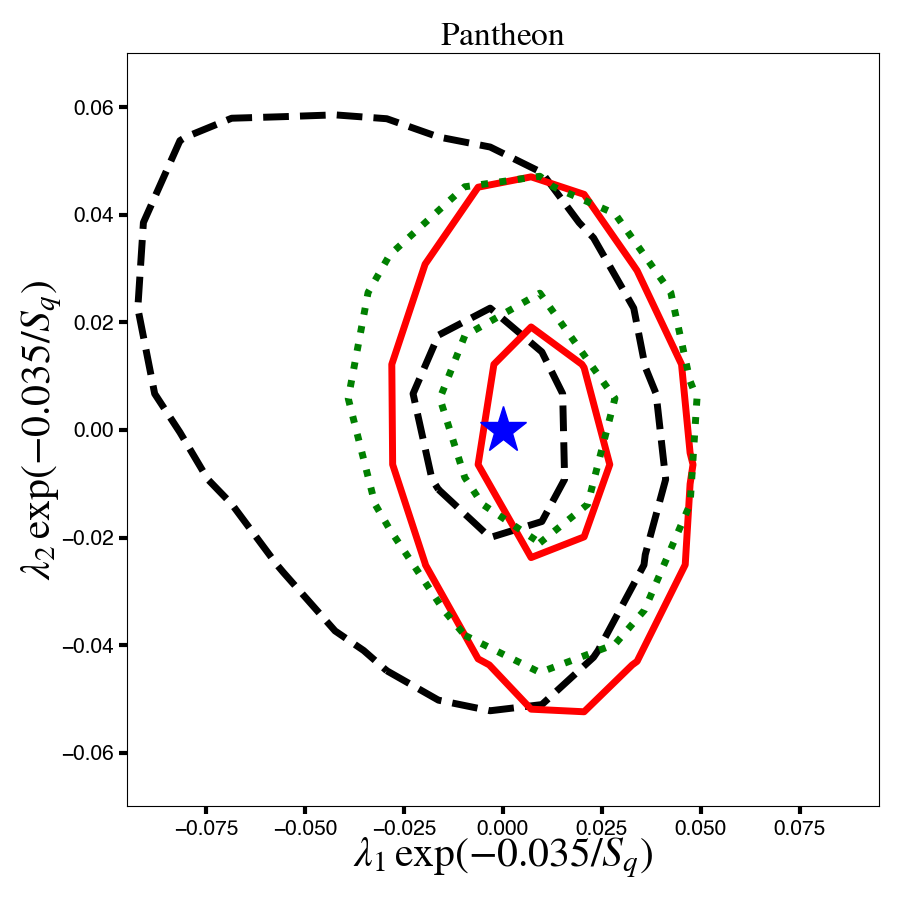}
    \caption{Constraints on the eigenvalues of the quadrupole of the Hubble parameter using the JLA (left) and Pantheon (right) datasets. The contours are obtained with the $\chi^2$ method with redshifts in the CMB frame (solid red), heliocentric frame (dashed black) and HD redshifts %which include corrections for the peculiar velocities 
    (dotted green). The contours show the 1- and 2-$\sigma$ limits. We find no significant evidence for deviation from isotropy (marked with the blue star) in any of the cases studied here.} 
    \label{fig:hubble_quadrupole_JLA}
\end{figure*}

\subsection{Datasets}\label{sec:data}

%We use both the JLA and Pantheon SN~Ia samples to constrain the multipolar contributions to the cosmological parameters.
%For our analysis 
We use the most recent {SN~Ia} lightcurve parameters and redshift  %magnitude-redshift relation 
data from the JLA \citep{2014A&A...568A..22B} and Pantheon \citep{2018ApJ...859..101S} compilations. Figure~\ref{fig:SNe_direc} shows the sky coverage of the two samples, with crosses showing Pantheon supernovae directions and circles showing JLA supernovae directions. All points are coloured according to the redshift of the supernova in the CMB frame, $z_{\rm CMB}$ (as defined below).

The cosmographic representation of the luminosity distance \eqref{eq:series} is {generically expected to be divergent for $z>1$ \citep{Cattoen:2007sk,Macpherson:2021gbh}.
However, the approximation of the Taylor series to the exact distance formula in isotropic cosmology is expected to be reasonable for redshifts close to 1 \citep[e.g., see][]{Arendse2020}}, at least for testing cosmologies close to the \lcdm\ model \citep{Aviles:2014rma}. 
Most 
Over 97\% of \sn\ in the JLA and Pantheon datasets %($  >$97\% for both compilations) 
have $z<1$, with the SN~Ia with the largest redshift has 
highest-redshift SN~Ia being $z = 1.3$ and $z = 2.3$, % $z = 2.26$, 
respectively.  Hence, for the majority of \sn, the Taylor series should provide a valid description of the distances. %} \ahr{check that this maximum redshift is correct.} \hmr{can we also maybe quote a percentage of how many SNe have z>1?}
%only valid out to $z=1$. 
%However, since we are interested in constraining anisotropies with an exponential decay scale, our constraints on the dipole and quadrupole components will be predominantly driven by the lowest-redshift \sn\ in the sample. 
We therefore adopt the cosmographic representation for \emph{all} \sn\ in both samples. 
{The anisotropic features that we are constraining are exponentially decaying with redshift and thus the main results of our analysis are predominantly determined by the lowest-redshift \sn\ in the sample.}
%, even for those with $z>1$.
%We note that while the cosmographic expression for the luminosity distance is typically valid only out to $z \sim 1$ since we are interested in the higher multipoles with an exponential decay scale. 
%Since the constraints for these higher-order terms are predominantly driven by the low-redshift \sn, the cosmographic expansion can be used as a valid description of the observables 
%\sdr{improve language}\hmr{I tried this - please change if the meaning is different!}
%\hmr{Add a sentence here noting the redshift range of the SNe and that some lie outside the radius of convergence of cosmography, but this is ok because ...}

Peculiar velocity (PV) corrections {based on estimates within the \lcdm\ model} are usually applied to the measured redshifts of nearby \sn\ in order to alleviate the motions of these \sn\ with respect to the CMB frame. There has been a recent debate in the literature about the consistency of these corrections and their impact on the evidence for cosmic acceleration \citep{Colin:2019,RH2020}. 
Therefore, we 
evaluate the impact of PV corrections on our constraints by presenting results inferred from three different redshift frames.
%Therefore, {we derive our parameter constraints three times, using different ways of assigning redshifts to the SNe~Ia. } %for all \aht{of our derived parameter} constraints we use three different redshifts for all SN~Ia in each sample to assess the impact of these corrections on our constraints. 
%Specifically, 
We consider: 
1) Heliocentric (Hel) redshifts{:} the measured redshifts of each {SN~Ia} in the heliocentric frame{;}
2) CMB-frame redshifts{:} the heliocentric redshifts corrected via a boost {of the Earth} to the CMB frame \citep[using the CMB dipole as inferred by][]{2020A&A...641A...1P}{;}
and 3) Hubble diagram (HD) redshifts{:}  %. The latter are 
CMB-frame redshifts \textit{with} PV corrections applied to individual \sn.  %, as is commonly used in isotropic studies \aht{of} the Hubble diagram. 
We will adopt the {CMB-frame redshifts} {in our fiducial analysis.} 
While redshifts in the heliocentric frame are not usually used for parametrising distances at cosmological scales, they are useful as a reference in model-independent analysis, e.g. in cases where we might not wish to assume that the dipole in the CMB is a purely observer-kinematic effect. 
As part of our analysis, we will also fit for the best-fit rest frame --- i.e., not \textit{a priori} constraining this to be the CMB frame --- for both samples of SNe~Ia.
%we in addition consider the best fit restframe with respect to the samples of SN~Iae. 

% as the fiducial case.
%\ahr{In the above our citation \citet{2020A&A...641A...1P} reads a bit weird. Should be cited as just 'Planck Collaboration'?}\hmr{I think you can change this in the bib file}

Previous studies %constraining the dipole in the deceleration parameter
using various {SN~Ia} compilations and {data reduction} %correction 
methods have reached differing conclusions about the significance of a dipole in the deceleration parameter. Some works have found no significant dipole and report consistency with {the} \lcdm\ {model}  \citep{2019PhRvL.122i1301S,2019MNRAS.486.5679Z,RH2020}, 
while others claim a deviation from isotropy at a level that challenges the use of the FLRW geometry at low redshift
\citep{2012JCAP...02..004C,Bengaly:2016,Colin:2018ghy}. 
Motivated by this discrepancy, we study of the impact of different %input corrections 
analysis assumptions
on the constraints of the dipole in the deceleration parameter.  
In particular, we test the impact of the PV corrections 
(through the use of the three different redshift frames outlined above)
on both datasets.
\begin{table}
    \centering
    \caption{Parameter priors used in the inference for each of the models tested in this work in both the JLA and Pantheon analyses.}
    \begin{tabular}{|c|c|c|}
    \hline 
    Parameter    &  Prior & Multipole model implemented in \\
    \hline
    $q_m$     & U[-4, 4]& Quadrupole and Dipole\\ 
    $j_0-\Omega_k$ & U[-10, 10] & {Quadrupole and} Dipole \\
    $q_d$ & U[-10, 10] & Dipole \\
    $S_d$ & U[0.01, 4] & Dipole \\
    $\lambda_1$ & U[-2, 2] & Quadrupole \\
    $\lambda_2$ & U[-2, 2] &  Quadrupole \\
    $S_q$ & U[0.01, 4] & Quadrupole \\
    \hline 
    \end{tabular}
    
    \label{tab:priors}
\end{table}
For the JLA dataset, we also analyse the role of the PV covariance matrix and distance bias corrections. 
The latter are applied to $m_B$ after the corrections to light-curve width, colour, and host galaxy mass, in order to account for systematics arising from survey selection criteria \citep[see][for more details on these corrections]{2014A&A...568A..22B,2018ApJ...859..101S}.
For the Pantheon data{set}, corrections for the width- and colour-luminosity relation and distance biases to the {SN~Ia} distances have been applied before the data was made public.
%and so to speed up our calculations we use the public data including the corrections.
Therefore, we only test 
%a limited number of corrections, namely peculiar velocities, 
the impact of PV corrections on results using the Pantheon sample. 
In Appendix~\ref{appx:pantheon_corr}, we {fit the nuisance parameters simultaneously with the cosmology and} show that we obtain similar constraints as in our main analysis of the Pantheon sample. %these do not impact our analysis.

\section{Results}\label{sec:results}
\begin{table*}
    \centering
    \caption{Summary of constraints on the isotropic deceleration and curvature minus jerk parameters $q_m$ and $j_0-\Omega_k$, the eigenvalues of the quadrupole in the Hubble parameter $\lam_1$ and $\lam_2$, and the exponential decay scale of the quadrupole, $S_q$. All results shown here are found using the $\chi^2$ method.}
    \begin{tabular}{|l|l|c|c|c|c|c|}
    \hline
    Dataset & Redshift  & $q_m$ & $j_0$ - $\Omega_{\rm K}$    & $\lambda_1$ & $\lambda_2$ & $S_q$   \\
    \hline
%    &&& JLA && \\
     JLA & CMB  & -0.316 $^{+0.115}_{-0.117}$ & -0.373 $^{+0.403}_{-0.49}$ & 0.005 $^{+0.017}_{-0.023}$ & 0.002 $^{+0.022}_{-0.017}$ & 0.974 $^{+0.98}_{-0.974}$\\
    JLA & HD & -0.392 $^{+0.122}_{-0.11}$ & -0.109 $^{+0.521}_{-0.494}$ & 0.003 $^{+0.012}_{-0.015}$ & 0.005 $^{+0.015}_{-0.012}$ & 1.258 $^{+0.901}_{-1.258}$
 \\
    JLA & Hel & -0.404 $^{+0.116}_{-0.113}$ & -0.115 $^{+0.47}_{-0.562}$ & 0.006 $^{+0.013}_{-0.016}$ & 0.001 $^{+0.013}_{-0.014}$ & 1.253 $^{+0.962}_{-1.252}$\\
    %&&& Pantheon && \\
    Pantheon & CMB  & -0.448 $^{+0.076}_{-0.081}$ & 0.264 $^{+0.289}_{-0.374}$ & 0.011 $^{+0.008}_{-0.01}$ & -0.003 $^{+0.01}_{-0.009}$ & 1.564 $^{+0.853}_{-1.554}$ \\%-0.4358 $\pm$ 0.073 & 3.89 $\pm$ 2.849 \\
    Pantheon & HD &  -0.481 $^{+0.078}_{-0.067}$ & 0.38 $^{+0.331}_{-0.335}$ & 0.072 $^{+1.552}_{-0.907}$ & -0.136 $^{+0.663}_{-1.816}$ & 0.002 $^{+0.001}_{-0.002}$ \\%-0.4814 $\pm$ 0.074 & 0.135 $\pm$ 0.671 \\ 
    Pantheon & Hel & -0.49 $^{+0.078}_{-0.073}$ & 0.408 $^{+0.283}_{-0.399}$ & -0.007 $^{+0.022}_{-0.019}$ & 0.003 $^{+0.015}_{-0.017}$ & 0.275 $^{+1.096}_{-0.265}$ \\%-0.4466 $\pm$ 0.0736 & -4.934 $\pm$ 2.19 \\
    
    \hline
    \end{tabular}
    
    \label{tab:hq_pantheon}
\end{table*}

We present our inferred constraints on the quadrupole of the Hubble parameter in Section~\ref{ssec:quad}, and on the dipole of the deceleration parameter %including different common corrections 
in Section~\ref{ssec:dip}. %We note that 
We constrain these %higher-order multipoles 
independently, % in this analysis, 
i.e., when constraining the quadrupole, we set the dipole term to zero, and vice versa. We perform Bayesian analysis based on the constrained $\chi^2$ metod for both the JLA and the Pantheon sample of \sn , and consider an independent frequentist MLE analysis for the JLA sample. The priors that we use for each model parameter in the Bayesian analyses %in all fits 
are summarised in Table~\ref{tab:priors}. %\ahr{We are referring to table 2 before table 1. Switch around? }
\begin{figure*}
\centering
\includegraphics[width=.48\textwidth, trim = 0 20 0 20]{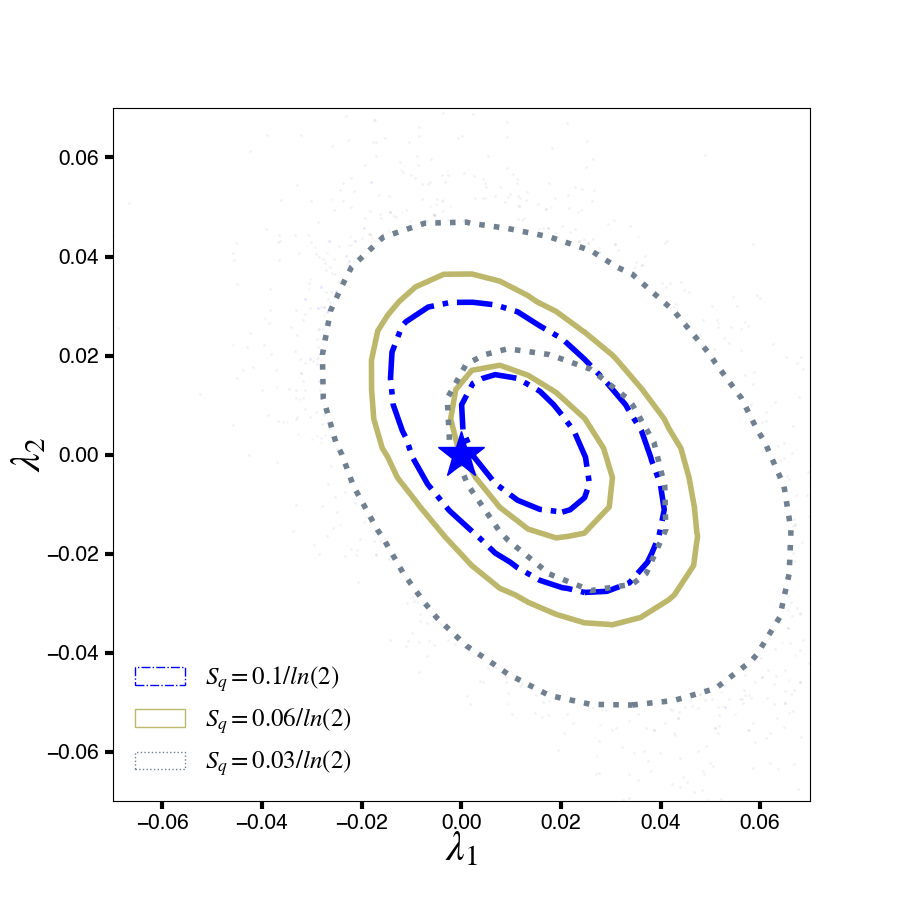}
\includegraphics[width=.48\textwidth, trim = 0 20 0 20]{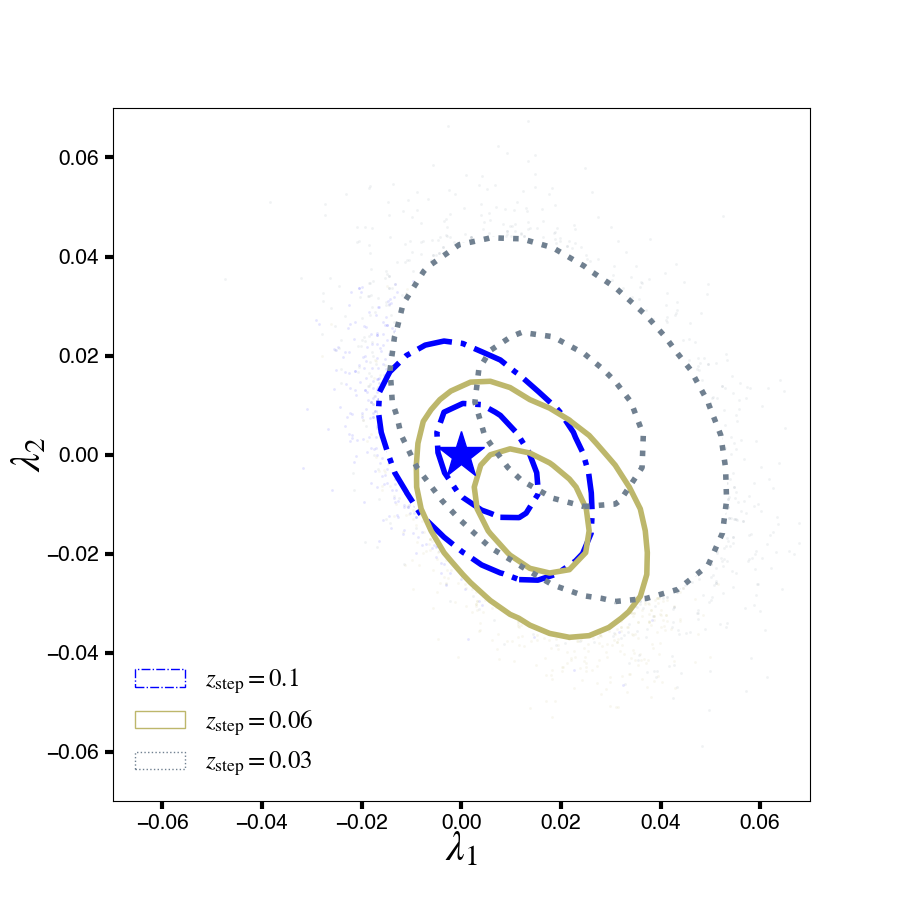}
\caption{Constraints on the eigenvalues of the quadrupole $\lam_1$, $\lam_2$ for the parametrisations with a fixed scale (left) and fixed redshift step value (right). The scale values are varied between $\frac{0.03}{ln(2)}, \frac{0.06}{ln(2)}$, and $\frac{0.1}{ln(2)}$ and the redshift steps at $0.03, 0.06$, and 0.1. As expected, the constraints are worsened for small step values since there are fewer SNe~Ia in the redshift range. All cases are consistent with isotropy. }
\label{fig:alt_quad_models}
\end{figure*}

\subsection{Constraints on the quadrupole}
\label{ssec:quad}

For the constraints on the quadrupole we use the exponential decay model for $\mathcal{F}(z, S_q)$ as the fiducial case with the scale as a free parameter. We also evaluate the constraints with the scale parameter fixed as well as %. Moreover, we test %a slightly different parametrisation of $\mathcal{F}$ as 
a step function in redshift, as described in Section~\ref{sec:cosmography}.

\citet{2013Ap&SS.343..747P} constrained the dipole, quadrupole, and octupole moments of the bulk motion of a set of nearby galaxies. 
%We fix the eigendirections of the quadrupole using the results of \citet{2013Ap&SS.343..747P}, 
We fix the eigendirections of the quadrupole in the Hubble parameter to coincide with their best-fit results of % by setting %$(l,b) =$ 
{$\boldsymbol{\theta}_1$=(118,85)\textdegree, $\boldsymbol{\theta}_2$=(341,4)\textdegree, and $\boldsymbol{\theta}_3$=(71,-4)\textdegree\ in galactic angular coordinates $(l, b)$.
%in galactic angular coordinates $(l, b)$, where $b=0$ correspond to the galactic disk and $l=b=0$ is the direction of the center of the Milky Way.} %in \eqref{eq:h_z_quadrupole}. 
%HM: I removed this explanation because it sounds like this is standard for these coordinates?
With these eigendirections, we then constrain the eigenvalues of the quadrupole, $\lambda_1$ and $\lambda_2$, and {its} decay scale $S_q$, {along with the monpolar parameters $q_m$ and $j_0-\Omega_K$ of the analysis. }%and the monopole of the deceleration and curvature minus jerk parameters. 
%For the fiducial case, w
{W}e have repeated the {analysis} allowing the eigendirections $\boldsymbol{\theta}_1,\boldsymbol{\theta}_2,$ and $\boldsymbol{\theta}_3$ to vary, and {have found no significant improvements in the profile likelihood for any alternative eigenbasis. } %find that all directions we tested give consistent results.

Figure~\ref{fig:hubble_quadrupole_JLA} shows our constraints on {the quadrupole in the Hubble parameter as obtained} %$\lambda_1$, $\lambda_2$, and $S_q$ 
from the JLA {and the Pantheon datasets} using the $\chi^2$ method. 
We show the amplitude of the quadrupole contribution at redshift $z=0.035$ (or on scales of $\sim 100 h^{-1}$ Mpc), namely%\footnote{\aht{Due to the strong degeneracies of $\lambda_1$ and $\lambda_2$ with $S_q$, combining the parameters in this way gives the most intuitive representation of the amplitude. 
%constraining the amplitude parameters directly give the most accurate quantification of the quadrupole on the given scale.}} 
$\lambda_1 \,{\rm exp}(-0.035/S_q)$ and $\lambda_2\, {\rm exp}(-0.035/S_q)$. %, which allows us to gain more physical intuition into the size of the quadrupole than assessing these quantities alone. 
Dashed black contours show constraints using the heliocentric redshifts, solid red contours show those using CMB-frame redshifts, and dotted green contours show those using HD redshifts. Our results are consistent with zero in all cases and show no significant change between redshift frames. 

In Table~\ref{tab:hq_pantheon} we summarise our constraints on all parameters for both the JLA and Pantheon datasets obtained with the $\chi^2$ method. 
We show constraints using heliocentric, CMB-frame, and HD redshifts for both datasets.
For all cases, we find results consistent with $\lambda_1=\lambda_2=0$ 
at the $\sim 1 \sigma$ level. 
From the 95$\%$ confidence level in Figure~\ref{fig:hubble_quadrupole_JLA} and using \eqref{eq:ampcalc}, we place an upper limit on the total quadrupole amplitude of $\sim 10\%$ at scales of $\sim 100\,h^{-1}$\,Mpc (or $z = 0.035$).  %$\lesssim 4\%$ 
%$\lambda_1 <2.9\%$ and $\lambda_2 < 4.2 \%$ at $\sim 100\,h^{-1}$\,Mpc (or $z= 0.035$), and total quadrupole of 5$\%$ which can be seen from Figure~\ref{fig:hubble_quadrupole_JLA}
Therefore, the few-percent quadrupole %amplitude 
predicted in \citep{Macpherson:2021gbh} is consistent with current data. In Section~\ref{sec:forecast}, we forecast improvements on our constraint for upcoming low-redshift surveys such as the Zwicky Transient Facility \citep[ZTF;][]{2021arXiv211007256D} or the Young Supernova Experiment \citep[YSE;][]{2021ApJ...908..143J}.

\begin{table*}%[h!]
    \centering%
    \caption{Constraints on the isotropic deceleration and curvature minus jerk parameters $q_m$ and $j_0-\Omega_k$, the eigenvalues and exponential decay of the quadrupole in the effective Hubble parameter, $\lam_1, \lam_2$, and $S_q$. Results here are found using the MLE method and the JLA SN~Ia dataset. The $m_B$ bias corrections are removed and $\sigma_z$ is set to zero. The p-value in the right-most column is the probability of the null hypothesis (isotropic universe model) relative to the model with a non-zero quadrupole.}
%   \resizebox{\textwidth}{!}{%
    %\begin{adjustbox}{width=\textwidth,center}
    \begin{tabular}{|l|l|c|c|c|c|c|c|}%
    \hline 
    Redshift & $q_m$ & $j_0-\Omega_{\rm K}$ & $\lambda_1$ & $\lambda_2$ & $S_q$ & p-value \\
    \hline
    CMB &  -0.160  &  -0.455 & 0.109 & -0.0396 & 0.0110 & 0.67  \\
     % CMB  & \\%-0.4358 $\pm$ 0.073 & 3.89 $\pm$ 2.849 \\
    HD & -0.260  & -0.159  & 4.78 & -4.27 & 0.0028 &  0.67 \\%-0.4814 $\pm$ 0.074 & 0.135 $\pm$ 0.671 \\ 
    Hel & -0.151 & -0.496  & -0.00713 & 0.0095 & 24.8 &  0.81\\
    \hline
    \end{tabular}% 
    %} $-7.13\times10^{-3}$ $9.50\times10^{-3}$
    \label{tab:MLE_hq_pantheon}
\end{table*}
Table~\ref{tab:MLE_hq_pantheon} shows our constraints on the quadrupole parameters using the MLE method for all three redshift frames. 
For all cases, our result{s are} consistent with isotropy (zero quadrupole) {at the $\sim 1 \sigma$ level}, which can be seen {from the p-value for the isotropic null hypothesis as quoted} in the right-most column {of the table.} %which shows the probability of the isotropic model, relative to the quadrupole model, being high.

%In addition to the fiducial case above, 
We also test two different parametrisations of the quadrupole %which  
{that determine} %essentially fix 
the %parameters that determine the 
redshift region where the quadrupole dominates. First, we fix the exponential decay scale to $S_q = 0.03/{\rm ln}(2), 0.06/{\rm ln}(2)$, and $0.1/{\rm ln}(2)$. These choices imply ${\rm exp}(-z/S_q)=1/2$ for redshifts $z=0.03,0.06$, and 0.1, respectively. 
Second, we treat the quadrupole as a step function in redshift, i.e.
we set 
%it to zero beyond a set of chosen redshift values 
$\mathcal{F}(z\leq z_{\rm step}{, z_{\rm step}})=0$ and $\mathcal{F}(z>z_{\rm step}{,z_{\rm step}})=1$ for $z_{\rm step} = 0.03, 0.06$, and 0.1. %We use three well-motivated values to 
%We fix the exponential decay scale to $S_q = 0.03/{\rm ln}(2), 0.06/{\rm ln}(2), 0.1/{\rm ln}(2)$ and the step function value to $z_{\rm step} = 0.03, 0.06, 0.1$. 
%We select these values to be in the low-$z$ (i.e. $z \leq 0.1$) regime while still being sufficiently higher than the minimum redshift in the SN~Ia compilations.
These redshift values %\sout{are well-motivated since they} 
all lie in the low-$z$ (i.e. $z \leq 0.1$) regime --- where we expect the anisotropy to be strongest --- while still being sufficiently above the minimum redshift in the SN~Ia compilations. %\sout{--- so as to maintain a good level of constraining power with sufficient \sn.} 
The left panel of Figure~\ref{fig:alt_quad_models} shows the posterior distribution for the eigenvalues {of the quadrupole}, using the $\chi^2$ method, for the three %cases of 
{exponential decay profiles} %fixed $S_q$ 
for the Pantheon sample. The right panel of Figure~\ref{fig:alt_quad_models} shows the same constraints for the three cases of the step function. We find similar constraints for the both the %model with the 
fixed redshift step and the exponential decay model with the fixed decay scale. 
For all cases shown here we use the CMB frame redshifts, however, we find similar results for the Helicontric and HD frame redshifts, all indicating a quadrupole feature consistent with zero at the $\sim 1 \sigma$ level. %the same level of consistency between frames as for the fiducial case in Figure~\ref{fig:hubble_quadrupole_JLA}.
%The resulting posterior distribution for the eigenvalues is shown in Figure~\ref{fig:alt_quad_models}. 
In all of these cases, we thus find no significant deviation from the isotropic null hypothesis. 
We also find that performing the same fits with the JLA sample gives consistent results.

\begin{figure*}
    \centering
    \includegraphics[width=.49\textwidth, trim = 10 0 10 60]{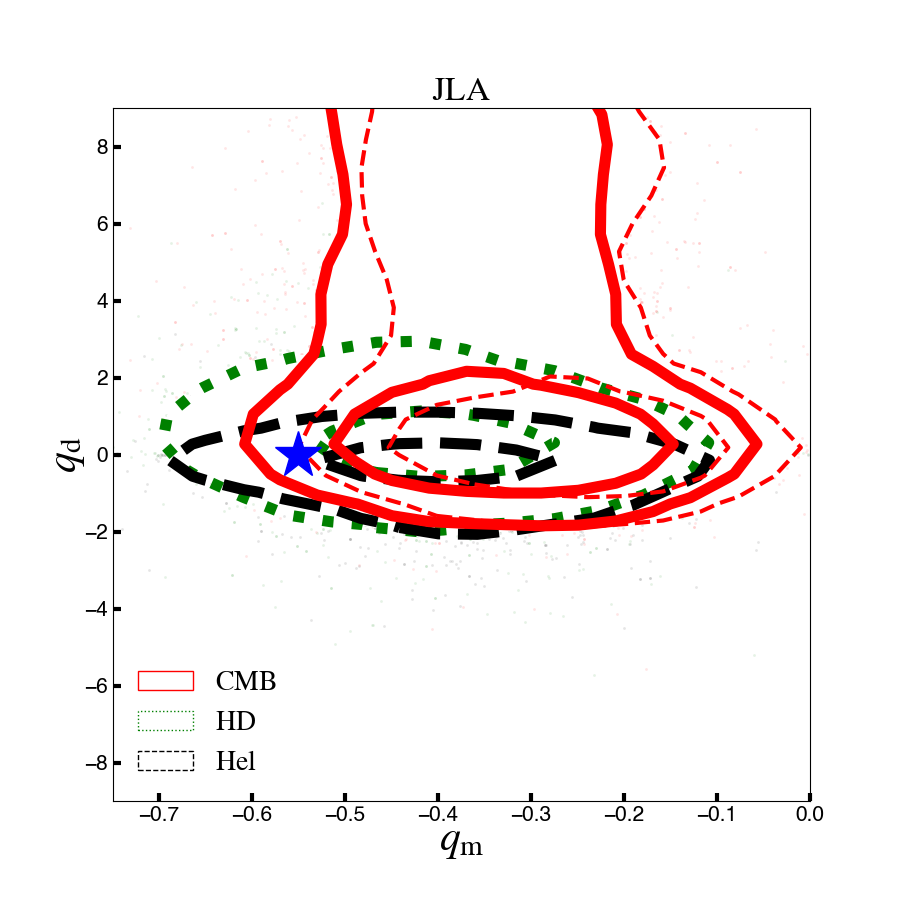}
        \includegraphics[width=.49\textwidth, trim = 10 0 10 60]{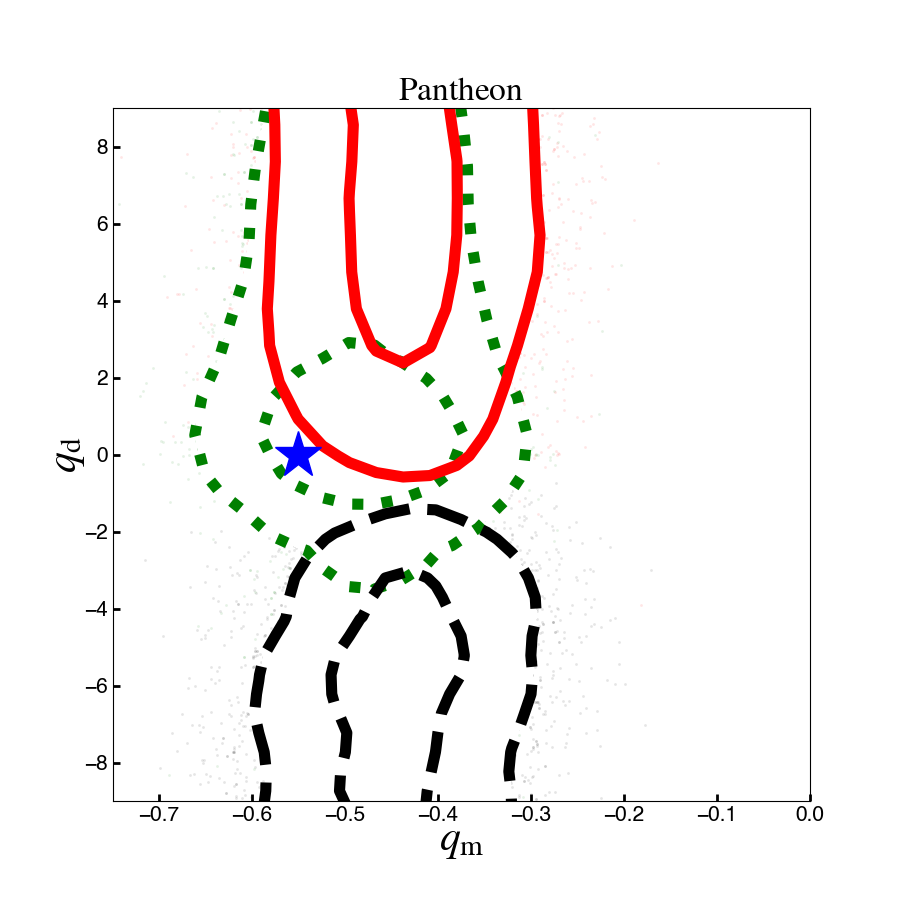}

    \caption{Left panel: Constraints on the monopole and dipole terms of the deceleration parameter using the JLA compilation. The constraints are shown for the heliocentric (thick dashed black), CMB frame redshift (solid red) and the HD redshifts (dotted green). All contours contain the PV covariance matrix. Thin dashed red contours in the left panel show the CMB frame constraints without the distance bias corrections. Right panel: The same constraints for the Pantheon compilation. All contours represent the 1- and 2-$\sigma$ constraints.}
    \label{fig:decel_dipole_JLA}
\end{figure*}

%------------------------------------------------------
\subsection{Dipole of the deceleration parameter}
\label{ssec:dip}

%Previous studies using various SN~Ia compilations and correction methods have reached differing conclusions about the significance of a dipole in the deceleration parameter. Some works have found no significant dipole and report consistency with \lcdm\ \citep{2010MNRAS.401.1409C,2014MNRAS.439.1855H,2018MNRAS.478.5153S,2019PhRvL.122i1301S,2019MNRAS.486.5679Z,2020ApJ...894...68R,1966ApJ...143..379K}, while others claim a detection of deviation from isotropy \citep{2012JCAP...02..004C,Colin:2018ghy}. 
%HM: I moved this up to the new 'Datasets' subsection, where we discuss using different corrections

%HM: we now have this stated above in "datasets" subsection
%In this section, we constrain the dipole in the deceleration parameter using the JLA and Pantheon SN~1a datasets. We analyse the impact of different corrections to data in the analysis on inferred values of the amplitude and decay scale of the dipole. Specifically, we vary the statistical model, distance bias corrections, peculiar velocity covariances, as well as directly comparing the JLA and Pantheon compilations for identical cases.

%In our analysis 
{In our main analysis, } we {set} %fix 
the direction of the dipole {in the effective deceleration parameter} to coincide with the CMB dipole {as found by} \citet{2020A&A...641A...1P}, namely $(l, b)$ = (264.021, 48.523)\textdegree. 
{In order to ensure that the CMB dipole direction is indeed an optimal direction for the dipolar signature, } %However, similar to \citet{Colin:2018ghy},
we test for the best-fit direction by varying the dipole direction and comparing the likelihood of the fit for different directions on the sky (see Appendix~\ref{appx:dip_direc}). Using the MLE method, we find {the direction that optimises the profile Likelihood function to} %the maximal likelihood direction to 
closely coincide with the direction of the CMB dipole{, %similarly to what was 
as was also found in \citet{Colin:2018ghy}}. %, so we present only those results here. 

% -----------------
% 1. dipole JLA, Pantheon using chisq
% -----------------
The left panel of Figure~\ref{fig:decel_dipole_JLA} shows our constraint contours in the $q_d$--$q_m$ plane for the JLA dataset using the $\chi^2$ method, including the PV covariance matrix. The right panel shows the same constraints for the Pantheon dataset. 
%In both panels, red contours show the results from CMB-frame redshifts, green contours show HD redshifts, and black contours show heliocentric redshifts. 
In both panels, solid red contours show the results from CMB-frame redshifts, dotted green contours show HD redshifts, and thick dashed black contours show heliocentric redshifts. 
All constraints include the distance bias corrections, with the exception of the thin dashed red contours in the left panel, which show the CMB-frame constraints for JLA with these corrections removed. 
%We also assess the impact of distance bias corrections on our results. Dashed red contours in the left panel of Figure~\ref{fig:decel_dipole_JLA} show the constraints for JLA with the distance bias corrections removed (to be compared with the equivalent case including the corrections in solid red).
%We find that 
Removing these corrections %{from the constrained $\chi^2$ results} 
does not significantly impact our constraints, and so we retain them for the rest of our analysis. 
 
We summarise our constraints on the deceleration parameter using the $\chi^2$ method in both the JLA and Pantheon data in Table~\ref{tab:cases_tab}. We show constraints on the monopole $q_m$, the dipole amplitude $q_d$, the decay scale $S_d$, and the (isotropic) jerk minus curvature {parameter} $j_0-\Omega_K$. We show all three redshift cases {with PV covariance included in the estimated errors,} as well as the CMB and heliocentric redshifts %\sout{with and} 
without PV covariance {contributions} \citep[see][for details on the components of the covariance matrix]{2014A&A...568A..22B}.
%\sout{The fiducial case includes all components of the covariance matrix of errors \citep[see][for details]{2014A&A...568A..22B}, whereas ``without PV cov" excludes the covariance matrix pertaining to the error in the PV corrections.} 
For the JLA dataset, we find $q_d = 2.18 ^{+3.353}_{-2.724}$ for the CMB frame redshifts, and when adding the PV corrections we find $q_d =  -0.004 ^{+1.08}_{-0.785}$. In all but one case, the JLA dataset yields a dipole consistent with zero. In the case of JLA heliocentric redshifts without the PV covariance matrix, we find a significant dipole at the $ 3.3\sigma$ level. For the rest of this work, we compute the significance of our results as %the significance of the dipole in each case as 
$\sqrt{2}$ times in the inverse error function of the $n^{\rm th}$ percentile that is consistent with isotropy, i.e. $q_d = 0$. 
For Pantheon, both heliocentric and CMB-frame redshifts yield a dipole at $3.43\sigma$ and $2.17\sigma$ significance, respectively (including the PV covariance matrix). After applying the PV corrections (i.e. using HD redshifts), the dipole is consistent with zero within $1\sigma$ for both samples.

Comparing the constraints from the JLA and Pantheon compilations in the left and right panels of Figure~\ref{fig:decel_dipole_JLA}, respectively,  %(Figures~\ref{fig:decel_dipole_JLA} and~\ref{fig:q0_dipole_Pantheon}) 
we find that the posterior distributions 
%have a fairly similar shape %is consistent 
are similar for the CMB and HD redshifts, with the $1 \sigma$ contours of the %JLA and Pantheon 
two samples (close to) overlapping.
For redshifts in the heliocentric frame, we find {an overlap of the 2 $\sigma$ contours (but not the $1 \sigma$ contours)}, which {indicates a moderate}  inconsistency between the two samples.  %at the $\lesssim 2 \sigma$ level. 
We note that there have been several updates between the two compilations, e.g. the redshift measurements for a subsample, additional objects at high-$z$ photometric calibration, retraining of the lightcurve fitting method. 
We remade 
the constraints in Figure~\ref{fig:decel_dipole_JLA} %peated this fit 
%We tested the impact of 
using \textit{only} the SNe~Ia in common between the two compilations as well as using the same redshift measurement reported for $z_{\rm hel}$ --- i.e., by using $z_{\rm hel}$ reported in one sample for \textit{all} objects with the respective magnitudes from each sample, and vice versa. However, in both of these tests we still find an inconsistency between the samples at the $\lesssim 2 \sigma$ level. 
We
therefore %, while we 
cannot %directly 
attribute this inconsistency to the difference in objects between the samples or in %the change in the 
any difference in reported redshifts. The {source of the systematic differences pointed out here are important to clarify and} should be further investigated with larger, improved samples such as {the} Pantheon+ {compilation}
%which are expected imminently 
\citep{brout22:P+}. %\hmt{This would be interesting to investigate with the improved Pantheon+ sample \hmr{suggest possible reasons? cite Pantheon+?}.}
\begin{table*}
    \centering
    \caption{The median and 68$\%$ C.L. constraints on the monopole and dipole moments of the deceleration parameter for both JLA and Pantheon compilations. Here we compute the parameters for the CMB frame, HD (see text for details) and heliocentric frame redshifts. We also evaluate the parameters with and without the covariance matrix for the peculiar velocity corrections (for the CMB and heliocentric frames) for the JLA compilation. } %\sdr{moved MLE to separate table}}
    \begin{tabular}{|c|c|c|c|c|c|c|}
    \hline
     Dataset & Covariance & Redshift & $q_m$ & $j_0$ - $\Omega_{\rm K}$ & $q_d$  & $S_d$\\
     \hline
%         &&& JLA && \\
%    \hline

    JLA & With PV cov  & CMB & -0.348 $^{+0.128}_{-0.107}$ & -0.28 $^{+0.404}_{-0.558}$ & 1.016 $^{+3.262}_{-1.561}$ & 0.143 $^{+0.006}_{-0.006}$ \\
    JLA & With PV cov  & HD &   -0.413 $^{+0.124}_{-0.119}$ & -0.048 $^{+0.446}_{-0.623}$ & 0.034 $^{+0.737}_{-0.322}$ & 0.141 $^{+0.007}_{-0.006}$\\
    JLA & With PV cov  & Hel & -0.399 $^{+0.115}_{-0.113}$ & -0.129 $^{+0.511}_{-0.501}$ & -0.066 $^{+0.167}_{-0.447}$ & 0.142 $^{+0.006}_{-0.006}$\\
    %Without PV cov & HD & & &  \\
    JLA & Without PV cov & CMB &-0.343 $^{+0.103}_{-0.122}$ & -0.296 $^{+0.405}_{-0.553}$ & 2.379 $^{+2.868}_{-2.609}$ & 0.026 $^{+0.035}_{-0.026}$\\
    JLA & Without PV cov & Hel & -0.315 $^{+0.121}_{-0.104}$ & -0.380 $^{+0.449}_{-0.465}$ & -6.806 $^{+1.087}_{-3.189}$ & 0.028 $^{+0.008}_{-0.009}$\\
%    \hline
%    &&& Pantheon & & \\
%    \hline
    Pantheon & With PV cov & CMB & -0.439 $^{+0.076}_{-0.073}$ & 0.240 $^{+0.325}_{-0.323}$ & 5.414 $^{+4.486}_{-1.705}$ & 0.020 $^{+0.007}_{-0.009}$ \\
    Pantheon & With PV cov & HD  &  -0.481 $^{+0.073}_{-0.071}$ & 0.373 $^{+0.295}_{-0.374}$ & 0.696 $^{+4.002}_{-1.19}$ & 0.021 $^{+0.034}_{-0.021}$\\
    Pantheon & With PV cov & Hel & -0.445 $^{+0.076}_{-0.078}$ & 0.252 $^{+0.303}_{-0.365}$ & -6.001 $^{+2.037}_{-3.111}$ & 0.027 $^{+0.007}_{-0.01}$ \\
        \hline
    \end{tabular}
    
    \label{tab:cases_tab}
    
\end{table*}

\begin{figure}
    \centering
    \includegraphics[width=.46\textwidth, trim = 0 10 40 10]{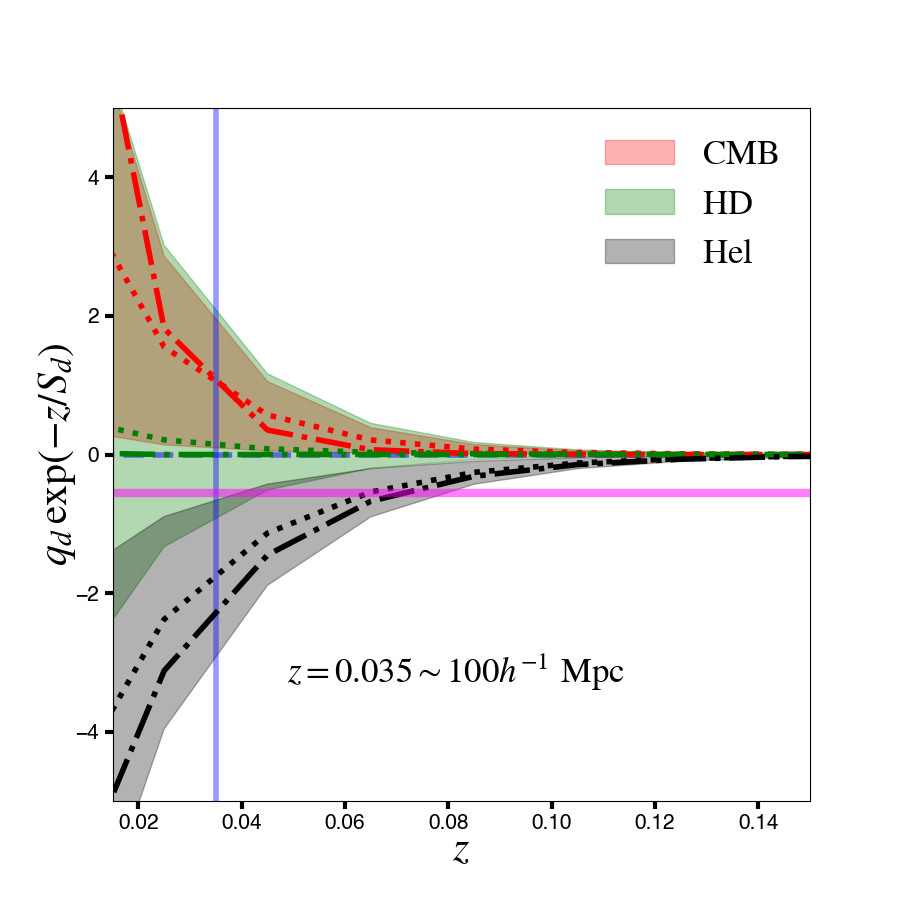}
    \caption{The evolution of the deceleration parameter dipole with redshift. We show the dipole amplitude as a function of redshift in the CMB (red), HD (green) and heliocentric (black) frames. The solid lines are the inferred values from the $\chi^2$ method applied to the JLA data (without PV covariance matrix for a direct comparison with the MLE method), whereas the dotted lines are for the $\chi^2$ method applied to the Pantheon data. The dashdot lines are the result from the MLE method applied to the JLA data. Shaded regions show the 2-$\sigma$ bounds for the $\chi^2$ constraints. The shaded region are the 2-$\sigma$ bounds for the constraints from the $\chi^2$ method for the Pantheon compilation. The magenta line shows the magnitude of the monopole in the standard cosmological model, for comparison.   %We see from this plot that the CMB frame marginally (at the $\sim 2 \sigma$ or lower level) suggests a non-zero dipole whereas there is no suggestion of a dipole when using the HD redshifts. We see a significant dipole when using the heliocentric redshifts.  The x-axis is truncated to the region where the dipole amplitude is significant. %HM: moved these sentences to the text as they are discussing the figure
    }
    \label{fig:dipole_summary}
\end{figure}

% -----------------
% 2. dipole JLA using MLE
% -----------------

{For the Pantheon dataset, the reported magnitudes have already been calibrated for}  %Since we are only fitting the $m_B$ which has already been corrected for the 
stretch, colour, and host galaxy properties of the \sn\ , within a cosmological model. We can {therefore} only use the constrained $\chi^2$ model for the Pantheon dataset. In Appendix~\ref{appx:pantheon_corr}, we test the impact of these {magnitude calibrations} %corrections 
on the cosmolog{ical constraints}, {by repeating the analysis} using the light curve %\sout{fit} 
parameters provided with the Pantheon compilation. We find %a negligible 
no correlation between the {SN~Ia} standardisation and {the} cosmological parameters {of our analysis}. %, as shown in Figure~\ref{fig:nuisance_params}. 
%We therefore find that 
{O}ur results with the corrected $m_B$ Pantheon data in the main analysis are thus recovered within the more model-independent approach examined in Appendix~\ref{appx:pantheon_corr}.  %robust for cosmological inference 
%\aht{within the distance parametrisation of this analysis}. 

 \begin{table}  
    \centering
    \caption{Constraints on the isotropic deceleration and curvature minus jerk parameters $q_m$ and $j_0-\Omega_k$, and the magnitude and exponential decay scale of the dipole in the effective deceleration parameter, $q_d$ and $S_d$. Results here are obtained with the MLE method and the JLA SN~Ia data set. The $m_B$ bias corrections are removed and $\sigma_z$ is set to zero. The p-value in the right-most column is the probability of the null hypothesis (isotropic universe model) relative to the model with a non-zero dipole.}
 %\resizebox{.5\textwidth}{!}{%
     \begin{tabular}{|c|c|c|c|c|c|}
    \hline
    Redshift & $q_m$ & $j_0$ - $\Omega_{\rm K}$ & $q_d$  & $S_d$ & p-value \\
     \hline
    CMB & -0.174 & -0.416 & 14.1 & 0.0122 & 0.024 \\
    HD & -0.256 & -0.174 & 10.4  & 0.00084 & 0.67 \\
    Hel & -0.158 & -0.488 & -8.13 & 0.0261 & $7.9\times10^{-5}$  \\
    \hline
    \end{tabular}%
 %   }
    \label{tab:cases_tab_mle}
    
\end{table}

Table~\ref{tab:cases_tab_mle} shows our constraints on the deceleration parameter for the JLA dataset using the MLE method. Again, we consider all three redshift cases.
%both the HD and heliocentric redshift cases. 
For the HD redshifts, the dipole signal is consistent with zero. For the heliocentric redshifts, we find a significant dipole with {best fit values} $q_d=-8.13$ and $S_d = 0.0261$ {and p-value $= 7.9\times10^{-5}$}. 
This result is consistent with the equivalent case in Table~\ref{tab:cases_tab} (heliocentric redshifts without PV cov) using the $\chi^2$ method.\footnote{We note that the constrained $\chi^2$ results in Table~\ref{tab:cases_tab} contain the distance bias corrections, whereas the MLE results in Table~\ref{tab:cases_tab_mle} \textit{do not} contain them. 
However, the addition of bias corrections makes little difference on our results for both statistical methods, and we therefore may still safely compare results between statistical methods. }  %\ahr{I moved text from below up to a footnote here.}
In the CMB frame, we find a preferred dipole with opposite sign of that in the heliocentric frame, albeit the significance of the signature is lowered. The change of sign of the preferred dipole in the deceleration parameter is due to the partial degeneracy between this dipole and the special-relativistic boost of the observer \citep[see Section~5 of][]{Heinesen:2020bej}. This result differs from that of the analogous analysis %is inconsistent with the equivalent case
using the $\chi^2$ method, for which we {found} no significant dipole signature.
%\ahr{Put in the exact results of Antonin here. Comment that this does not agree with the JLA CMB frame results of Table~\ref{tab:cases_tab}. }
% Suhail is checking the chisq results (JLA, CMB, without PV cov) which don't agree with the MLE CMB result for JLA

So far in our analysis, we have maintained the velocity of the observer to coincide with the best-fit velocity as inferred from the dipole in the CMB.
{If the dipole anisotropy in SN~Ia data is purely due to our kinematic motion and the CMB dipole is of purely kinematic origin as well, we should infer a similar observer velocity to that obtained from the CMB. 
%from SN~Ia data. 
We now }
leave %the magnitude of the velocity of the observer 
the amplitude of the observer velocity as a free parameter in an isotropic analysis, maintaining the direction fixed to that of the CMB dipole. %, and infer the best-fit velocity. 
We have repeated the analysis, allowing%also allowed 
the direction to vary, and find the maximum likelihood direction to closely coincide with that of the CMB dipole (see Appendix~\ref{appx:dip_direc}). 
For this test, %\sout{we use redshifts in the heliocentric frame} 
{we neglect} the PV covariance contributions to the total error covariance matrix. {Inclusion of PV covariance increases the error bars by $\sim 20\%$, but gives overall similar results to those quoted below.}
For the JLA \sn, we find a velocity $v = 258.15 ^{+57.9}_{-61.2}$ km/s {relative to the heliocentric frame} using the constrained $\chi^2$ method %\hmr{add JLA for chisq without PV cov, but check with PV cov too}
%MLE analysis for the JLA dataset (keeping the direction fixed to that of the CMB dipole), we find a best-fit value of 
and $v = 252$~km/s using the MLE method (with a p-value of $0.018$). Both of these velocities are consistent with the recent result in \cite{2021arXiv211103055H},
however, both are discrepant from that inferred from the CMB dipole \citep[$369.82 \pm 0.11$ km/s;][]{2020A&A...641A...1P}. %, according to our analysis. 
Using the $\chi^2$ method for the Pantheon \sn, we find a best-fit velocity of 240 $^{+57.0}_{-36.2}$ km/s, which is in agreement with our other results. 
{This suggests an additional contribution to the dipole in \sn\ data beyond that of a special-relativistic boost of the observer to the rest-frame of the CMB.}

%\hmr{clarify that this following statement is for a dipole analysis}

{We found a significant dipole in the deceleration parameter %$q_d=-8.13$ 
using the MLE method for the case of JLA \sn\ in the heliocentric frame (see Table~\ref{tab:cases_tab_mle}), with an observer velocity coinciding with that inferred from the CMB dipole. }
Instead keeping the observer velocity as a free parameter in this anisotropic analysis --- i.e., allowing both the kinematic and geometric dipoles to be free parameters --- removes the significance of $q_d \neq 0$. %with p-value X \hmr{Asta: add the p-value here}.
%Further, we find that %the inclusion of the observer velocity as a free parameter 
%allowing the observer velocity to vary removes the significance of the dipole in the deceleration parameter. 
Thus, we find that the dipole in the deceleration parameter is consistent with zero only if the frame of reference
%in a reference frame without PV corrections only if that reference frame 
is \emph{different} to the rest-frame of the CMB.
%Therefore, the price we must pay to remove the dipole in the deceleration parameter (without PV corrections to individual \sn) is to introduce a reference frame of the observer which is \textit{different} to that of the rest-frame of the CMB.
%an unmotivated frame of reference of the observer. \hmr{Asta: I don't really understand this previous sentence, can you try re-wording (or explaining via Slack)?}
% note this fit above is the isotropic case with an obs boost in direction of CMB dipole with free velocity
However, the HD frame results in Table~\ref{tab:cases_tab_mle} also show an insignificant dipole in the deceleration parameter. Thus, we conclude that the SN~Ia PV estimates in standard analyses \textit{can} account for the dipole in the deceleration parameter that we find here. 
Peculiar flows are indeed expected to give rise to anisotropies in the Hubble law of the type investigated in this paper, as we comment on in the discussion.

%This result is in contrast to that of \citet{Colin:2019}, who found the monopole of the deceleration parameter to be consistent with zero when allowing for a dipole component.} \hmr{Hayley todo: Comment on the differences in methods}

\begin{figure}
    \centering
    \includegraphics[width=.48\textwidth]{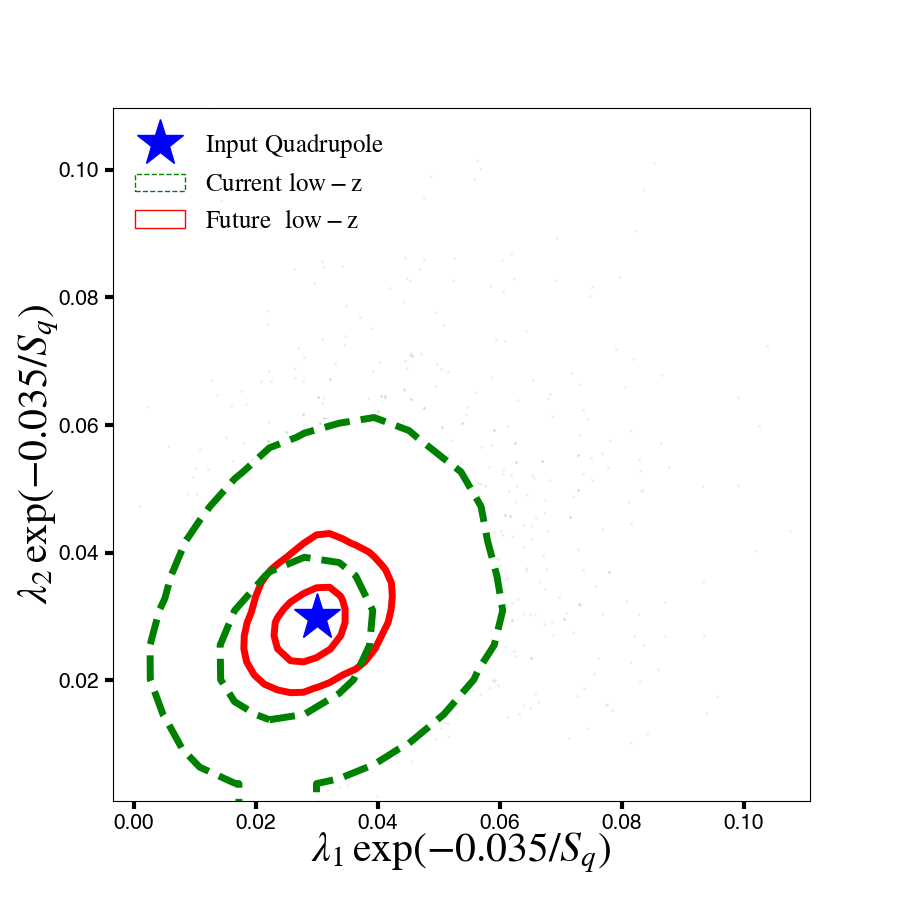}
    \caption{Posterior distribution of $\lambda_1 \cdot {\rm exp} (-z /S_q)$, $\lambda_2 \cdot {\rm exp} (-z /S_q)$ {for the artifical input model with a 2.9\% quadrupole at $z = 0.035$, corresponding to $\sim$100 $h^{-1}$\,Mpc scales.
    %the $\sim$~100~$h^{-1}$~Mpc scale. 
    %The posterior is evaluated} at $z = 0.035$, corresponding to scales of 100 $h^{-1}$\,Mpc \citep{Macpherson:2021gbh}. 
    %\ahr{I don't think that we need a citation to our paper here?  }
    %The forecasts are computed assuming two cases, an 
    The forecasts for a SN~Ia compilation of the same size as the current Pantheon compilation is shown as green contours, and for a compilation with 5 $\times$ larger low-$z$ anchor samples is shown as red contours. Contours show the 1- and 2-$\sigma$ regions, respectively. We mark the input quadrupole value in our forecasts with a blue star.}}
    \label{fig:nonzero_quad}
\end{figure}

In Figure~\ref{fig:dipole_summary} we show the exponentially-decaying dipole amplitude as a function of redshift for the different statistical methods and datasets used here. Different colours represent the three redshift frames we use, as indicated in the legend. Solid lines show best-fit values obtained using the $\chi^2$ method with JLA \sn, dotted lines show $\chi^2$ {best-}fit {values for} Pantheon \sn, and dot-dashed lines show results using the MLE method with JLA \sn. Shaded regions show the 2-$\sigma$ bounds for the $\chi^2$ constraints. The horizontal magenta line shows the magnitude of the monopole, for comparison, and the vertical blue line {marks the scale} %shows redshift of
$z=0.035$ {corresponding to a distance scale} of $\sim 100 h^{-1}$ Mpc. 
%We see from this plot that 
The Pantheon data using the CMB frame redshifts marginally %(at the $\lesssim 2 \sigma$ level) 
suggests a non-zero dipole at the $\sim 2 \sigma$ level, whereas we find no suggestion of a dipole when using the HD redshifts. We find a significant dipole in both datasets when using the heliocentric redshifts. This figure is a summary of our main results, while illustrating the redshift ranges for which a non-zero dipole (with this parametrisation) might be important. %However, we still require improved constraints to determine its amplitude more concretely. 

%we show the difference when removing the bias constraints on \hmr{one specific case}, which we can clearly see makes little difference to the contours. 
% HM: here fig:TBA could be a contour plot with one dashed contour showing an equivalent result without the bias corrections? we can re-word the prev sentence as necessary then.

Crucially, we find that in both left and right panels of Figure~\ref{fig:decel_dipole_JLA}, the posterior distribution of the monopole $q_m$ is not significantly correlated with the value of the dipole $q_d$. Hence, the assumption on the value of $q_d =0$ in the isotropic cosmography does not significantly impact the inferred $q_m$. 
Further, from both Figure~\ref{fig:decel_dipole_JLA} and Table~\ref{tab:cases_tab}, we can see that the boost to the CMB frame, and the PV corrections, do not significantly impact the inferred value of the monopole, $q_m$, when using the $\chi^2$ method. %For the dipole, however, the CMB frame transformation changes the sign and amplitude of the best fit $q_d$ value. 
Inferences of $q_0$ using isotropic cosmography in the literature \citep[e.g.][]{Bernal2016,2019MNRAS.483.4803L,2019PhRvL.122f1105F} are consistent with the $q_m$ value we find with the $\chi^2$ method at the 1--2$\sigma$ level.
Our results using the MLE method also show minimal change in the value of the monopole $q_m$ with redshift frame (see Table~\ref{tab:cases_tab_mle}). However, the values of the monopole in the heliocentric and CMB frames are $q_m=-0.158$ and $q_m=-0.174$, respectively, which deviate from the value within \lcdm\ of $q_0\approx -0.55$. %We note that 
The likely cause of this difference between the %$q_m$ inference from the 
two methods %presented here 
is the assumption of the redshift evolution of the population of \sn\ lightcurve width ($x_1$) and colour ($c$) parameters. The $\chi^2$ method accounts for survey selection as a function of redshift whereas the MLE method assumes no redshift dependence {in the distributions of the intrinsic supernova parameters}. Since the SN~Ia surveys are impacted by Malmquist bias, i.e.{,} they preferentially detect brighter SNe~Ia at higher redshifts, {the failure to account for such bias, or doing so in an incorrect manner,} %not sampling the underlying magnitude distribution or accounting for the selection bias 
can impact the value of the monopole term $q_m$. 
Such an impact has been recently discussed %extensively 
in the literature \citet{Colin:2019,RH2020}. %Since our work focuses on the higher order multipoles and not the monopole, we do not further discuss this and defer to the previous studies in the literature.
Our findings agree with both \citet{Colin:2019} and \citet{RH2020} for the relevant statistical method, and therefore further investigation into the {appropriate way of accounting for} %need to account for 
survey selection as a function of redshift is necessary to clarify this debate. 
%\hmr{Can this be attributed to using only low-redshift SNe?}}

%\hmr{Can we comment on $q_m=-0.158$ in the heliocentric and CMB cases for MLE? This seems significant and is quite far off -0.5.}
%In summary, we find that our constraints on the monopole contribution to the deceleration parameter are robust to changes in statistical method, redshift frame, and PV corrections, and between the JLA and Pantheon datasets.

\subsection{Forecast of constraints on the quadrupole in the Hubble parameter}\label{sec:forecast}

Ongoing and future surveys will discover a large number of \sn. %Type Ia supernovae.  Surveys such as %the Zwicky Transient Facility \citep[ZTF;][]{2021arXiv211007256D} and the Young Supernova Experiment \citep[YSE;][]{2021ApJ...908..143J} 
ZTF and YSE will increase the low-redshift {SN~Ia} sample %manifold with significantly improved 
and significantly improve systematic errors. For {regional anisotropies that decay towards larger scales,} %the exponentially decaying models, 
improvements in low-redshift data will make the most difference to our constraining power. In this section, we forecast the constraints on the quadrupole in the Hubble parameter from the improved low-$z$ samples. We start with a simulated realisation with uncertainties corresponding to the current Pantheon compilation, and then increase the number of low-$z$ samples to coincide with expected future datasets.

%We present a ``mock" future sample compared to the constraints presented from current data.

We infer distances to \sn\ for an input model with a quadrupole in the Hubble parameter such that $\lambda_1 \cdot {\rm exp} (-z /S_q) = \lambda_2 \cdot {\rm exp} (-z /S_q) = 0.029$ in \eqref{eq:h_z_quadrupole} at $z = 0.035$.
%5\%$. 
%The amplitude of the quadrupole is motivated by testing a theoretically predicted scenario from \citet{Macpherson:2021gbh} {\bf does this make sense}.
This {induced} $\sim 7\%$ quadrupole amplitude is motivated both by the upper limit on the quadrupole we find here 
%found in the present analysis %results we find here 
as well as {the numerical results obtained by} \citet{Macpherson:2021gbh}. In the latter, the authors found a quadrupole in the Hubble parameter of a few percent on $\sim 100 h^{-1}$ Mpc scales in a set of general-relativistic cosmological simulations.
We take the redshift and distance modulus error distribution {of the mock \sn} to be the same as the Pantheon data. We augment the low-$z$ ($z \leq 0.1$) anchor sample to five times its size such that the total number of low-$z$ \sn\ is 1055. This sample size is conservatively well within the limit of data already obtained by current and ongoing low-$z$ {SN~Ia} surveys. For comparison, we also use simulated distances for a low-$z$ anchor sample of the same size as the current Pantheon compilation. 
%\hmr{Not sure if this should go here but: is there some way we can put a limit on the minimum \% quadrupole we can detect with the samples we use? i.e., can we translate our null-detection into an upper limit somehow?}

Figure~\ref{fig:nonzero_quad} shows our forecast constraints for the input cosmology with a non-zero quadrupole for a future low-$z$ survey (solid red contours) and for a sample consistent with current low-$z$ catalogues (dashed green contours). The blue star represents the 
%eigenvalues $\lambda_1$ and $\lambda_2$ 
values of the input cosmology.
We find that the improved low-$z$ anchor sample will be able to detect a 7\% quadrupole at 100~$h^{-1}$~Mpc scales with 5$\sigma$ significance. 

\section{Discussion and Conclusion}
\label{sec:disc}

The assumption of isotropy is a central feature of the standard cosmological model and %therefore needs to 
must be empirically tested. 
Any universe with structure will necessarily contain anisotropies in the {$d_L$--$z$} relation. Therefore --- once the data is precise enough to resolve such anisotropies ---, 
anisotropic contributions to low-redshift data need to be included for a realistic cosmological fit. Such anisotropies could also impact local distance determinations, e.g.{,} those based on Cepheids  \citep{Riess2022:h0} or TRGB \citep{Freedman2021}, in the case of an anisotropic distribution of \sn.
In this work, we have presented the first constraints on the theoretically motivated quadrupole moment of the effective Hubble parameter {in the distance-redshift relation. Th{e} quadrupole {moment} physically arises from the anisotropic expansion of space around the observer {as incorporated in the shear tensor}. Using the SN~Ia magnitude-redshift relation, we find no significant %deviation from isotropy of 
quadrupole in the effective Hubble parameter, % in our analysis, 
with our best-fit quadrupole amplitude being consistent with zero.
%We find no significant deviation of the %current constraints on the 
%eigenvalues of the quadrupole $\lambda_1$ and $\lambda_2$ {from zero, which are the expected values within} an isotropic universe. 
This constraint holds for both the JLA and Pantheon compilations, and is robust to the changes in redshift frames and likelihood methods considered here. Our results are unchanged when including \lcdm\ modelled corrections for peculiar motions of the \sn\ with respect to the CMB 
frame.

We {have} placed an upper bound of  
%2.9\% and 4.2\% (95\% C.L.) for each of the eigenvalues $\lambda_1$, $\lambda_2$ at 100 $h^{-1}$ Mpc within our exponential decay model, corresponding to a 
$\sim 10\%$ {for the} quadrupolar anisotropy within our exponential decay model (using Eq.~\ref{eq:ampcalc}). Since the decay scale can be degenerate with the value of the eigenvectors, we also constrain the parametrisations with a fixed decay scale, as well as a fixed step value in redshift, and find results consistent with the fiducial model. 
%Our analysis indicates that the \emph{magnitude} of anisotropies in the \sn\ distance-redshift relation} might be too small to resolve the current tension in the Hubble parameter. %HM: I removed this because I don't think this is true now we have correctly calculated the amplitude to be 10% max :) 
%However, more precise studies with upcoming larger datasets must be done in order to cement this result. 
We stress again that this anisotropic contribution to the Hubble parameter is not degenerate with the SN~Ia absolute luminosity, unlike the monopole $H_m$.
It can therefore be constrained %\emph{entirely} 
by the magnitude-redshift relation of \sn\ without needing external calibrators to first constrain the SN~Ia absolute magnitude prior to the cosmological fit.

%Studies in the literature, 
\citet{2013Ap&SS.343..747P} used %galaxy catalogs to constrain the local quadrupole. They use 
the Revised Flat Galaxy Catalogue (RFGC) to constrain the quadrupole (shear) at the 100~$h^{-1}$~Mpc scale, finding
%Our constraints might be compared directly with the results for the 
eigenvalues of ${\lambda}_1 = 7.27 {\%} \pm 1.54 {\%}$ and ${\lambda}_2 = -2.43 {\%} \pm 1.46 {\%}$.
Since we use the same eigendirections as the RFGC study, we can %directly 
compare our constraints to their results. The RFGC quadrupole was found to be almost constant over the 80--170~$h^{-1}$~Mpc scales which they considered, therefore, our quadrupole constraints from the %$z_{\rm step}$ 
step function parametrisation (with $z_{\rm step}=0.03$) are the closest in formalism to the RFGC study. {We find agreement between our results and the RFGC measurements at the $\sim$2$\sigma$ level (but not at the 1$\sigma$ level)}.  %The RFGC measurements are significant at the $\sim 4 -5 \sigma$ level and thus disagree with our quadrupole eigenvalues --- which we find to be consistent with zero within our error bars.
%While we find that quadrupole is consistent with zero with the errors, the RFGC measurements are more significant, at the $\sim 1.5 - 4.5 \sigma$ level. 
It is %therefore 
important to %perform checks 
constrain this quadrupolar anisotropy with future larger datasets, as well as with independent probes.

We have also forecasted the precision of quadrupole measurements from future low-$z$ SN~Ia surveys. 
This forecast is timely, since the number of \sn\ available for cosmological studies {will} increas{e} manyfold {within this decade}. 
With {up}coming samples of SN{e}~Ia in the nearby Hubble flow, e.g., from ZTF or YSE, we can significantly improve {the} constraints {on the quadrupole moment of the Hubble parameter}. 
For our input signal we took %the lowest upper limit, i.e. of 
a quadrupole with $\sim 7\%$ amplitude %from $\lambda_1$ 
to test whether it can feasibly be constrained with future surveys. Specifically, we forecast that with $1055$ \sn\ we will have the potential to detect this quadrupole at $5\sigma$ significance. A sample of this size is also interesting {for} constrain{ing} the kinematic nature of the CMB dipole \citep[e.g.][]{2021arXiv211103055H}. Hence, the improve{d} low-$z$ {data} will be important for tests of the cosmic rest frame. 

We have also presented constraints on the dipole in the deceleration parameter. We focused on the impact of the statistical method as well as input data assumptions. 
We find that for the JLA compilation, %independent of the redshift frame used, 
the dipole is consistent with zero {at the $1\sigma$ level} when inferred using the $\chi^2$ method for all but one case. The only instance of a significant dipole occurs in the heliocentric frame without applying the PV covariance matrix. 
With the same inference method, we find that the Pantheon compilation indicates {marginal significance of a dipole} %consistency with isotropy
at the $\sim 2 \sigma$ level when using the CMB frame, however, {this dipolar signature vanishes} %this %is reduced to $\sim 0.01\sigma$ 
when applying the PV corrections to the SN~Ia redshifts. We note that for the MLE method, we similarly find that the CMB frame redshifts with PV corrections are consistent with isotropy. However when PV corrections are not applied, we find a significant dipole in both the CMB and heliocentric frame. 
In Figure~\ref{fig:dipole_summary}, we {have} present{ed} a summary of the dipole amplitude for the exponentially decaying case, illustrating its dependence on redshift for both statistical methods and datasets used here, as well as all redshift frames.

Recent improvements in the treatment of PV corrections have been shown to have a small impact on parameter constraints in isotropic cosmologies \citep[e.g. for $H_0$ in][]{2021arXiv211003487P}. In addition, \citet{2021arXiv210812497R} used an improved flow model to correct for PVs and found no evidence for departures from isotropy {once the PV corrections were applied}, consistent with our {findings.}  %results. % presented here.
 %We note that 
The theoretical framework developed by \citet{Heinesen:2020bej} %and employed in our analysis 
in principle allows us to account for anisotropic expansion of space and to infer peculiar velocities around a background model for both the observer and the sources. This will be a possibility with future low-$z$ SN~Ia samples that have significantly increased statistics.
%\aht{G}iven the marginal consistency of the Pantheon compilation with the assumption of isotropy when using the CMB frame redshifts, 
{It will be interesting to further tighten the constraints on anisotropies %found in this paper 
we find here using %with } %this will be an interesting investigation 
upcoming improve{d} SN~Ia magnitude-redshift data \citep[e.g.][]{brout22:P+}.

\section*{Acknowledgements}

We thank Ariel Goobar, Thomas Buchert, Dillon Brout, and Dan Scolnic for their valuable comments. SD acknowledges support from the Marie Curie Individual Fellowship under grant ID 890695 and a junior research fellowship at Lucy Cavendish College. HJM appreciates support received from the Herchel Smith Postdoctoral Fellowship Fund. 
AH acknowledges funding from the European Research Council (ERC) under the European Union's Horizon 2020 research and innovation programme (grant agreement ERC advanced grant 740021--ARTHUS, PI: Thomas Buchert). 
AB conducted an internship within ERC-adG ArthUs related to this work: \cite{Borderies:2021}. 
%%%%%%%%%%%%%%%%%%%%%%%%%%%%%%%%%%%%%%%%%%%%%%%%%%
\section*{Data Availability}

The data for the analysis is public and the software used is entirely based on publicly available packages and will be made available upon request.



\bibliographystyle{mnras}
\bibliography{anisotropy} % if your bibtex file is called example.bib

\begin{thebibliography}{}
\makeatletter
\relax
\def\mn@urlcharsother{\let\do\@makeother \do\$\do\&\do\#\do\^\do\_\do\%\do\~}
\def\mn@doi{\begingroup\mn@urlcharsother \@ifnextchar [ {\mn@doi@}
  {\mn@doi@[]}}
\def\mn@doi@[#1]#2{\def\@tempa{#1}\ifx\@tempa\@empty \href
  {http://dx.doi.org/#2} {doi:#2}\else \href {http://dx.doi.org/#2} {#1}\fi
  \endgroup}
\def\mn@eprint#1#2{\mn@eprint@#1:#2::\@nil}
\def\mn@eprint@arXiv#1{\href {http://arxiv.org/abs/#1} {{\tt arXiv:#1}}}
\def\mn@eprint@dblp#1{\href {http://dblp.uni-trier.de/rec/bibtex/#1.xml}
  {dblp:#1}}
\def\mn@eprint@#1:#2:#3:#4\@nil{\def\@tempa {#1}\def\@tempb {#2}\def\@tempc
  {#3}\ifx \@tempc \@empty \let \@tempc \@tempb \let \@tempb \@tempa \fi \ifx
  \@tempb \@empty \def\@tempb {arXiv}\fi \@ifundefined
  {mn@eprint@\@tempb}{\@tempb:\@tempc}{\expandafter \expandafter \csname
  mn@eprint@\@tempb\endcsname \expandafter{\@tempc}}}

\bibitem[\protect\citeauthoryear{{Andrade}, {Bengaly}, {Santos}  \&
  {Alcaniz}}{{Andrade} et~al.}{2018}]{Andrade:2018}
{Andrade} U.,  {Bengaly} C. A.~P.,  {Santos} B.,   {Alcaniz} J.~S.,  2018,
  \mn@doi [\apj] {10.3847/1538-4357/aadb90}, \href
  {https://ui.adsabs.harvard.edu/abs/2018ApJ...865..119A} {865, 119}

\bibitem[\protect\citeauthoryear{{Arendse} et~al.,}{{Arendse}
  et~al.}{2020}]{Arendse2020}
{Arendse} N.,  et~al., 2020, \mn@doi [\aap] {10.1051/0004-6361/201936720},
  \href {https://ui.adsabs.harvard.edu/abs/2020A&A...639A..57A} {639, A57}

\bibitem[\protect\citeauthoryear{Aviles, Bravetti, Capozziello  \&
  Luongo}{Aviles et~al.}{2014}]{Aviles:2014rma}
Aviles A.,  Bravetti A.,  Capozziello S.,   Luongo O.,  2014, \mn@doi [Phys.
  Rev. D] {10.1103/PhysRevD.90.043531}, 90, 043531

\bibitem[\protect\citeauthoryear{{Bengaly}}{{Bengaly}}{2016}]{Bengaly:2016}
{Bengaly} C.~A.~P. J.,  2016, \mn@doi [\jcap] {10.1088/1475-7516/2016/04/036},
  \href {https://ui.adsabs.harvard.edu/abs/2016JCAP...04..036B} {2016, 036}

\bibitem[\protect\citeauthoryear{{Bengaly}, {Bernui}  \& {Alcaniz}}{{Bengaly}
  et~al.}{2015}]{Bengaly:2015}
{Bengaly} C.~A.~P. J.,  {Bernui} A.,   {Alcaniz} J.~S.,  2015, \mn@doi [\apj]
  {10.1088/0004-637X/808/1/39}, \href
  {https://ui.adsabs.harvard.edu/abs/2015ApJ...808...39B} {808, 39}

\bibitem[\protect\citeauthoryear{{Bernal}, {Verde}  \& {Riess}}{{Bernal}
  et~al.}{2016}]{Bernal2016}
{Bernal} J.~L.,  {Verde} L.,   {Riess} A.~G.,  2016, \mn@doi [\jcap]
  {10.1088/1475-7516/2016/10/019}, \href
  {http://adsabs.harvard.edu/abs/2016JCAP...10..019B} {10, 019}

\bibitem[\protect\citeauthoryear{{Betoule} et~al.,}{{Betoule}
  et~al.}{2014}]{2014A&A...568A..22B}
{Betoule} M.,  et~al., 2014, \mn@doi [\aap] {10.1051/0004-6361/201423413},
  \href {http://adsabs.harvard.edu/abs/2014A%26A...568A..22B} {568, A22}

\bibitem[\protect\citeauthoryear{Borderies}{Borderies}{2021}]{Borderies:2021}
Borderies A.,  2021, Inhomogeneities and Anisotropies in the Universe analyzed
  with Supernovae of Type Ia

\bibitem[\protect\citeauthoryear{{Brout} et~al.,}{{Brout}
  et~al.}{2022}]{brout22:P+}
{Brout} D.,  et~al., 2022, arXiv e-prints, \href
  {https://ui.adsabs.harvard.edu/abs/2022arXiv220204077B} {p. arXiv:2202.04077}

\bibitem[\protect\citeauthoryear{Buchert}{Buchert}{2000}]{Buchert:1999er}
Buchert T.,  2000, \mn@doi [Gen. Rel. Grav.] {10.1023/A:1001800617177}, 32, 105

\bibitem[\protect\citeauthoryear{Buchert, Ellis  \& van Elst}{Buchert
  et~al.}{2009}]{Buchert:2009wj}
Buchert T.,  Ellis G. F.~R.,   van Elst H.,  2009, \mn@doi [Gen. Rel. Grav.]
  {10.1007/s10714-009-0828-4}, 41, 2017

\bibitem[\protect\citeauthoryear{{Buchner} et~al.,}{{Buchner}
  et~al.}{2014}]{2014A&A...564A.125B}
{Buchner} J.,  et~al., 2014, \mn@doi [\aap] {10.1051/0004-6361/201322971},
  \href {http://adsabs.harvard.edu/abs/2014A%26A...564A.125B} {564, A125}

\bibitem[\protect\citeauthoryear{{Cai} \& {Tuo}}{{Cai} \&
  {Tuo}}{2012}]{2012JCAP...02..004C}
{Cai} R.-G.,  {Tuo} Z.-L.,  2012, \mn@doi [\jcap]
  {10.1088/1475-7516/2012/02/004}, \href
  {https://ui.adsabs.harvard.edu/abs/2012JCAP...02..004C} {2012, 004}

\bibitem[\protect\citeauthoryear{Cattoen \& Visser}{Cattoen \&
  Visser}{2007}]{Cattoen:2007sk}
Cattoen C.,  Visser M.,  2007, \mn@doi [Class. Quant. Grav.]
  {10.1088/0264-9381/24/23/018}, 24, 5985

\bibitem[\protect\citeauthoryear{Clarkson \& Umeh}{Clarkson \&
  Umeh}{2011}]{Clarkson:2011uk}
Clarkson C.,  Umeh O.,  2011, \mn@doi [Class. Quant. Grav.]
  {10.1088/0264-9381/28/16/164010}, 28, 164010

\bibitem[\protect\citeauthoryear{Clarkson, Ellis, Faltenbacher, Maartens, Umeh
  \& Uzan}{Clarkson et~al.}{2012}]{Clarkson:2011br}
Clarkson C.,  Ellis G. F.~R.,  Faltenbacher A.,  Maartens R.,  Umeh O.,   Uzan
  J.-P.,  2012, \mn@doi [Mon. Not. Roy. Astron. Soc.]
  {10.1111/j.1365-2966.2012.21750.x}, 426, 1121

\bibitem[\protect\citeauthoryear{{Colin}, {Mohayaee}, {Rameez}  \&
  {Sarkar}}{{Colin} et~al.}{2019a}]{Colin:2019}
{Colin} J.,  {Mohayaee} R.,  {Rameez} M.,   {Sarkar} S.,  2019a, arXiv
  e-prints, \href {https://ui.adsabs.harvard.edu/abs/2019arXiv191204257C} {p.
  arXiv:1912.04257}

\bibitem[\protect\citeauthoryear{Colin, Mohayaee, Rameez  \& Sarkar}{Colin
  et~al.}{2019b}]{Colin:2018ghy}
Colin J.,  Mohayaee R.,  Rameez M.,   Sarkar S.,  2019b, \mn@doi [Astron.
  Astrophys.] {10.1051/0004-6361/201936373}, 631, L13

\bibitem[\protect\citeauthoryear{Dam, Heinesen  \& Wiltshire}{Dam
  et~al.}{2017}]{Dam:2017xqs}
Dam L.~H.,  Heinesen A.,   Wiltshire D.~L.,  2017, \mn@doi [Mon. Not. Roy.
  Astron. Soc.] {10.1093/mnras/stx1858}, 472, 835

\bibitem[\protect\citeauthoryear{{Dhawan} et~al.,}{{Dhawan}
  et~al.}{2022}]{2021arXiv211007256D}
{Dhawan} S.,  et~al., 2022, \mn@doi [\mnras] {10.1093/mnras/stab3093}, \href
  {https://ui.adsabs.harvard.edu/abs/2022MNRAS.510.2228D} {510, 2228}

\bibitem[\protect\citeauthoryear{{Ellis}, {Nel}, {Maartens}, {Stoeger}  \&
  {Whitman}}{{Ellis} et~al.}{1985}]{1985PhR...124..315E}
{Ellis} G.~F.~R.,  {Nel} S.~D.,  {Maartens} R.,  {Stoeger} W.~R.,   {Whitman}
  A.~P.,  1985, \mn@doi [Phys. Rep.] {10.1016/0370-1573(85)90030-4}, \href
  {https://ui.adsabs.harvard.edu/abs/1985PhR...124..315E} {124, 315}

\bibitem[\protect\citeauthoryear{{Feeney}, {Peiris}, {Williamson}, {Nissanke},
  {Mortlock}, {Alsing}  \& {Scolnic}}{{Feeney}
  et~al.}{2019}]{2019PhRvL.122f1105F}
{Feeney} S.~M.,  {Peiris} H.~V.,  {Williamson} A.~R.,  {Nissanke} S.~M.,
  {Mortlock} D.~J.,  {Alsing} J.,   {Scolnic} D.,  2019, \mn@doi [\prl]
  {10.1103/PhysRevLett.122.061105}, \href
  {https://ui.adsabs.harvard.edu/abs/2019PhRvL.122f1105F} {122, 061105}

\bibitem[\protect\citeauthoryear{{Feroz}, {Hobson}  \& {Bridges}}{{Feroz}
  et~al.}{2009}]{2009MNRAS.398.1601F}
{Feroz} F.,  {Hobson} M.~P.,   {Bridges} M.,  2009, \mn@doi [\mnras]
  {10.1111/j.1365-2966.2009.14548.x}, \href
  {http://adsabs.harvard.edu/abs/2009MNRAS.398.1601F} {398, 1601}

\bibitem[\protect\citeauthoryear{{Freedman}}{{Freedman}}{2021}]{Freedman2021}
{Freedman} W.~L.,  2021, \mn@doi [\apj] {10.3847/1538-4357/ac0e95}, \href
  {https://ui.adsabs.harvard.edu/abs/2021ApJ...919...16F} {919, 16}

\bibitem[\protect\citeauthoryear{Heinesen}{Heinesen}{2020}]{Heinesen:2020bej}
Heinesen A.,  2020, \mn@doi [arXiv e-prints] {10.1088/1475-7516/2021/05/008},
  p. arxiv:2010.06534

\bibitem[\protect\citeauthoryear{Heinesen}{Heinesen}{2021}]{Heinesen:2021qnl}
Heinesen A.,  2021, \mn@doi [Phys. Rev. D] {10.1103/PhysRevD.104.123527}, 104,
  123527

\bibitem[\protect\citeauthoryear{{Heinesen} \& {Macpherson}}{{Heinesen} \&
  {Macpherson}}{2022}]{Heinesen:2021azp}
{Heinesen} A.,  {Macpherson} H.~J.,  2022, \mn@doi [\jcap]
  {10.1088/1475-7516/2022/03/057}, \href
  {https://ui.adsabs.harvard.edu/abs/2022JCAP...03..057H} {2022, 057}

\bibitem[\protect\citeauthoryear{Hogg, Eisenstein, Blanton, Bahcall, Brinkmann,
  Gunn  \& Schneider}{Hogg et~al.}{2005}]{Hogg:2004vw}
Hogg D.~W.,  Eisenstein D.~J.,  Blanton M.~R.,  Bahcall N.~A.,  Brinkmann J.,
  Gunn J.~E.,   Schneider D.~P.,  2005, \mn@doi [Astrophys. J.]
  {10.1086/429084}, 624, 54

\bibitem[\protect\citeauthoryear{{Horstmann}, {Pietschke}  \&
  {Schwarz}}{{Horstmann} et~al.}{2021}]{2021arXiv211103055H}
{Horstmann} N.,  {Pietschke} Y.,   {Schwarz} D.~J.,  2021, arXiv e-prints,
  \href {https://ui.adsabs.harvard.edu/abs/2021arXiv211103055H} {p.
  arXiv:2111.03055}

\bibitem[\protect\citeauthoryear{{Hutsem{\'e}kers}, {Cabanac}, {Lamy}  \&
  {Sluse}}{{Hutsem{\'e}kers} et~al.}{2005}]{Hutsemekers2005}
{Hutsem{\'e}kers} D.,  {Cabanac} R.,  {Lamy} H.,   {Sluse} D.,  2005, \mn@doi
  [\aap] {10.1051/0004-6361:20053337}, \href
  {https://ui.adsabs.harvard.edu/abs/2005A&A...441..915H} {441, 915}

\bibitem[\protect\citeauthoryear{{Hutsem{\'e}kers}, {Braibant}, {Pelgrims}  \&
  {Sluse}}{{Hutsem{\'e}kers} et~al.}{2014}]{Hutsemekers2014}
{Hutsem{\'e}kers} D.,  {Braibant} L.,  {Pelgrims} V.,   {Sluse} D.,  2014,
  \mn@doi [\aap] {10.1051/0004-6361/201424631}, \href
  {https://ui.adsabs.harvard.edu/abs/2014A&A...572A..18H} {572, A18}

\bibitem[\protect\citeauthoryear{{Jones} et~al.,}{{Jones}
  et~al.}{2021}]{2021ApJ...908..143J}
{Jones} D.~O.,  et~al., 2021, \mn@doi [\apj] {10.3847/1538-4357/abd7f5}, \href
  {https://ui.adsabs.harvard.edu/abs/2021ApJ...908..143J} {908, 143}

\bibitem[\protect\citeauthoryear{{Kalus}, {Schwarz}, {Seikel}  \&
  {Wiegand}}{{Kalus} et~al.}{2013}]{Kalus:2013}
{Kalus} B.,  {Schwarz} D.~J.,  {Seikel} M.,   {Wiegand} A.,  2013, \mn@doi
  [\aap] {10.1051/0004-6361/201220928}, \href
  {https://ui.adsabs.harvard.edu/abs/2013A&A...553A..56K} {553, A56}

\bibitem[\protect\citeauthoryear{{Kristian} \& {Sachs}}{{Kristian} \&
  {Sachs}}{1966}]{1966ApJ...143..379K}
{Kristian} J.,  {Sachs} R.~K.,  1966, \mn@doi [\apj] {10.1086/148522}, \href
  {https://ui.adsabs.harvard.edu/abs/1966ApJ...143..379K} {143, 379}

\bibitem[\protect\citeauthoryear{Laurent et~al.}{Laurent
  et~al.}{2016}]{Laurent:2016eqo}
Laurent P.,  et~al., 2016, \mn@doi [JCAP] {10.1088/1475-7516/2016/11/060}, 11,
  060

\bibitem[\protect\citeauthoryear{{Leibundgut} \& {Sullivan}}{{Leibundgut} \&
  {Sullivan}}{2018}]{Leibundgut2018}
{Leibundgut} B.,  {Sullivan} M.,  2018, \mn@doi [\ssr]
  {10.1007/s11214-018-0491-8}, \href
  {https://ui.adsabs.harvard.edu/abs/2018SSRv..214...57L} {214, 57}

\bibitem[\protect\citeauthoryear{{Lemos}, {Lee}, {Efstathiou}  \&
  {Gratton}}{{Lemos} et~al.}{2019}]{2019MNRAS.483.4803L}
{Lemos} P.,  {Lee} E.,  {Efstathiou} G.,   {Gratton} S.,  2019, \mn@doi
  [\mnras] {10.1093/mnras/sty3082}, \href
  {https://ui.adsabs.harvard.edu/abs/2019MNRAS.483.4803L} {483, 4803}

\bibitem[\protect\citeauthoryear{{MacCallum} \& {Ellis}}{{MacCallum} \&
  {Ellis}}{1970}]{1970CMaPh..19...31M}
{MacCallum} M.~A.~H.,  {Ellis} G.~F.~R.,  1970, \mn@doi [Comm. Math. Phys.]
  {10.1007/BF01645496}, \href
  {https://ui.adsabs.harvard.edu/abs/1970CMaPh..19...31M} {19, 31}

\bibitem[\protect\citeauthoryear{{Macaulay} et~al.,}{{Macaulay}
  et~al.}{2019}]{2019MNRAS.486.2184M}
{Macaulay} E.,  et~al., 2019, \mn@doi [\mnras] {10.1093/mnras/stz978}, \href
  {https://ui.adsabs.harvard.edu/abs/2019MNRAS.486.2184M} {486, 2184}

\bibitem[\protect\citeauthoryear{Macpherson \& Heinesen}{Macpherson \&
  Heinesen}{2021}]{Macpherson:2021gbh}
Macpherson H.~J.,  Heinesen A.,  2021, \mn@doi [Phys. Rev. D]
  {10.1103/PhysRevD.104.023525}, 104, 023525

\bibitem[\protect\citeauthoryear{Migkas, Pacaud, Schellenberger, Erler,
  Nguyen-Dang, Reiprich, Ramos-Ceja  \& Lovisari}{Migkas
  et~al.}{2021}]{Migkas:2021zdo}
Migkas K.,  Pacaud F.,  Schellenberger G.,  Erler J.,  Nguyen-Dang N.~T.,
  Reiprich T.~H.,  Ramos-Ceja M.~E.,   Lovisari L.,  2021, \mn@doi [Astron.
  Astrophys.] {10.1051/0004-6361/202140296}, 649, A151

\bibitem[\protect\citeauthoryear{Nielsen, Guffanti  \& Sarkar}{Nielsen
  et~al.}{2016}]{Nielsen:2015pga}
Nielsen J.~T.,  Guffanti A.,   Sarkar S.,  2016, \mn@doi [Sci. Rep.]
  {10.1038/srep35596}, 6, 35596

\bibitem[\protect\citeauthoryear{{Parnovsky} \& {Parnowski}}{{Parnovsky} \&
  {Parnowski}}{2013}]{2013Ap&SS.343..747P}
{Parnovsky} S.~L.,  {Parnowski} A.~S.,  2013, \mn@doi [\apss]
  {10.1007/s10509-012-1267-3}, \href
  {https://ui.adsabs.harvard.edu/abs/2013Ap&SS.343..747P} {343, 747}

\bibitem[\protect\citeauthoryear{Perivolaropoulos \& Skara}{Perivolaropoulos \&
  Skara}{2021}]{Perivolaropoulos:2021jda}
Perivolaropoulos L.,  Skara F.,  2021, arXiv, p. arXiv:2105.05208

\bibitem[\protect\citeauthoryear{{Peterson} et~al.,}{{Peterson}
  et~al.}{2021}]{2021arXiv211003487P}
{Peterson} E.~R.,  et~al., 2021, arXiv e-prints, \href
  {https://ui.adsabs.harvard.edu/abs/2021arXiv211003487P} {p. arXiv:2110.03487}

\bibitem[\protect\citeauthoryear{{Planck Collaboration}}{{Planck
  Collaboration}}{2020a}]{2020A&A...641A...1P}
{Planck Collaboration} 2020a, \mn@doi [\aap] {10.1051/0004-6361/201833880},
  \href {https://ui.adsabs.harvard.edu/abs/2020A&A...641A...1P} {641, A1}

\bibitem[\protect\citeauthoryear{{Planck Collaboration}}{{Planck
  Collaboration}}{2020b}]{2020A&A...641A...6P}
{Planck Collaboration} 2020b, \mn@doi [\aap] {10.1051/0004-6361/201833910},
  \href {https://ui.adsabs.harvard.edu/abs/2020A&A...641A...6P} {641, A6}

\bibitem[\protect\citeauthoryear{{Rahman}, {Trotta}, {Boruah}, {Hudson}  \&
  {van Dyk}}{{Rahman} et~al.}{2021}]{2021arXiv210812497R}
{Rahman} W.,  {Trotta} R.,  {Boruah} S.~S.,  {Hudson} M.~J.,   {van Dyk} D.~A.,
   2021, arXiv e-prints, \href
  {https://ui.adsabs.harvard.edu/abs/2021arXiv210812497R} {p. arXiv:2108.12497}

\bibitem[\protect\citeauthoryear{Rasanen}{Rasanen}{2009}]{Rasanen:2009mg}
Rasanen S.,  2009, \mn@doi [Phys. Rev. D] {10.1103/PhysRevD.79.123522}, 79,
  123522

\bibitem[\protect\citeauthoryear{Rasanen}{Rasanen}{2010}]{Rasanen:2010wz}
Rasanen S.,  2010, \mn@doi [Phys. Rev. D] {10.1103/PhysRevD.81.103512}, 81,
  103512

\bibitem[\protect\citeauthoryear{{Riess} et~al.,}{{Riess}
  et~al.}{2021}]{Riess2022:h0}
{Riess} A.~G.,  et~al., 2021, arXiv e-prints, \href
  {https://ui.adsabs.harvard.edu/abs/2021arXiv211204510R} {p. arXiv:2112.04510}

\bibitem[\protect\citeauthoryear{{Rubin} \& {Heitlauf}}{{Rubin} \&
  {Heitlauf}}{2020}]{RH2020}
{Rubin} D.,  {Heitlauf} J.,  2020, \mn@doi [\apj] {10.3847/1538-4357/ab7a16},
  \href {https://ui.adsabs.harvard.edu/abs/2020ApJ...894...68R} {894, 68}

\bibitem[\protect\citeauthoryear{{Saadeh}, {Feeney}, {Pontzen}, {Peiris}  \&
  {McEwen}}{{Saadeh} et~al.}{2016}]{2016PhRvL.117m1302S}
{Saadeh} D.,  {Feeney} S.~M.,  {Pontzen} A.,  {Peiris} H.~V.,   {McEwen} J.~D.,
   2016, \mn@doi [\prl] {10.1103/PhysRevLett.117.131302}, \href
  {https://ui.adsabs.harvard.edu/abs/2016PhRvL.117m1302S} {117, 131302}

\bibitem[\protect\citeauthoryear{{Scolnic} et~al.,}{{Scolnic}
  et~al.}{2018}]{2018ApJ...859..101S}
{Scolnic} D.~M.,  et~al., 2018, \mn@doi [\apj] {10.3847/1538-4357/aab9bb},
  \href {https://ui.adsabs.harvard.edu/abs/2018ApJ...859..101S} {859, 101}

\bibitem[\protect\citeauthoryear{{Scrimgeour} et~al.,}{{Scrimgeour}
  et~al.}{2012}]{2012MNRAS.425..116S}
{Scrimgeour} M.~I.,  et~al., 2012, \mn@doi [\mnras]
  {10.1111/j.1365-2966.2012.21402.x}, \href
  {https://ui.adsabs.harvard.edu/abs/2012MNRAS.425..116S} {425, 116}

\bibitem[\protect\citeauthoryear{{Secrest}, {von Hausegger}, {Rameez},
  {Mohayaee}, {Sarkar}  \& {Colin}}{{Secrest} et~al.}{2021}]{Secrest2021}
{Secrest} N.~J.,  {von Hausegger} S.,  {Rameez} M.,  {Mohayaee} R.,  {Sarkar}
  S.,   {Colin} J.,  2021, \mn@doi [\apjl] {10.3847/2041-8213/abdd40}, \href
  {https://ui.adsabs.harvard.edu/abs/2021ApJ...908L..51S} {908, L51}

\bibitem[\protect\citeauthoryear{{Soltis}, {Farahi}, {Huterer}  \&
  {Liberato}}{{Soltis} et~al.}{2019}]{2019PhRvL.122i1301S}
{Soltis} J.,  {Farahi} A.,  {Huterer} D.,   {Liberato} C.~M.,  2019, \mn@doi
  [\prl] {10.1103/PhysRevLett.122.091301}, \href
  {https://ui.adsabs.harvard.edu/abs/2019PhRvL.122i1301S} {122, 091301}

\bibitem[\protect\citeauthoryear{{Tripp}}{{Tripp}}{1998}]{1998A&A...331..815T}
{Tripp} R.,  1998, \aap, \href
  {https://ui.adsabs.harvard.edu/abs/1998A&A...331..815T} {331, 815}

\bibitem[\protect\citeauthoryear{Umeh}{Umeh}{2013}]{Umeh:2013UCT}
Umeh O.,  2013, PhD thesis, University of Cape Town, Faculty of Science,
  Department of Mathematics and Applied Mathematics

\bibitem[\protect\citeauthoryear{Visser}{Visser}{2004}]{Visser:2003vq}
Visser M.,  2004, \mn@doi [Class. Quant. Grav.] {10.1088/0264-9381/21/11/006},
  21, 2603

\bibitem[\protect\citeauthoryear{{Zhao}, {Zhou}  \& {Chang}}{{Zhao}
  et~al.}{2019}]{2019MNRAS.486.5679Z}
{Zhao} D.,  {Zhou} Y.,   {Chang} Z.,  2019, \mn@doi [\mnras]
  {10.1093/mnras/stz1259}, \href
  {https://ui.adsabs.harvard.edu/abs/2019MNRAS.486.5679Z} {486, 5679}

\makeatother
\end{thebibliography}

%\begin{thebibliography}{99}
%\end{thebibliography}

%%%%%%%%%%%%%%%%%%%%%%%%%%%%%%%%%%%%%%%%%%%%%%%%%%

%%%%%%%%%%%%%%%%% APPENDICES %%%%%%%%%%%%%%%%%%%%%

\appendix

\vspace{-0.3cm}
\section{Dependence on nuisance parameters}
\label{appx:pantheon_corr}
%We note that 
The JLA supernova compilation provides a catalog of peak apparent magnitude, $m_B$, light curve width, $x_1$, and colour, $c$, values along with the host galaxy masses. We therefore can marginalise over the nuisance parameters in the luminosity standardisation relation in \eqref{eq:obs_distmod}. For the Pantheon compilation, however, the publicly-available apparent luminosity has already been corrected for the width-luminosity and color-luminosity relations.
However, this fit was %for a cosmology that is 
explicitly based on the FLRW metric. We, therefore, verify the impact of these corrections by marginalising the SN~Ia width-luminosity, colour-luminosity and host galaxy mass distributions along with the cosmological parameters for our model. For this we use the publicly available lightcurve fit parameters and host galaxy masses for the Pantheon SNe~Ia.
\begin{figure*}[ht]
    \centering
    \includegraphics[width=.75\textwidth, trim = 0 20 0 20]{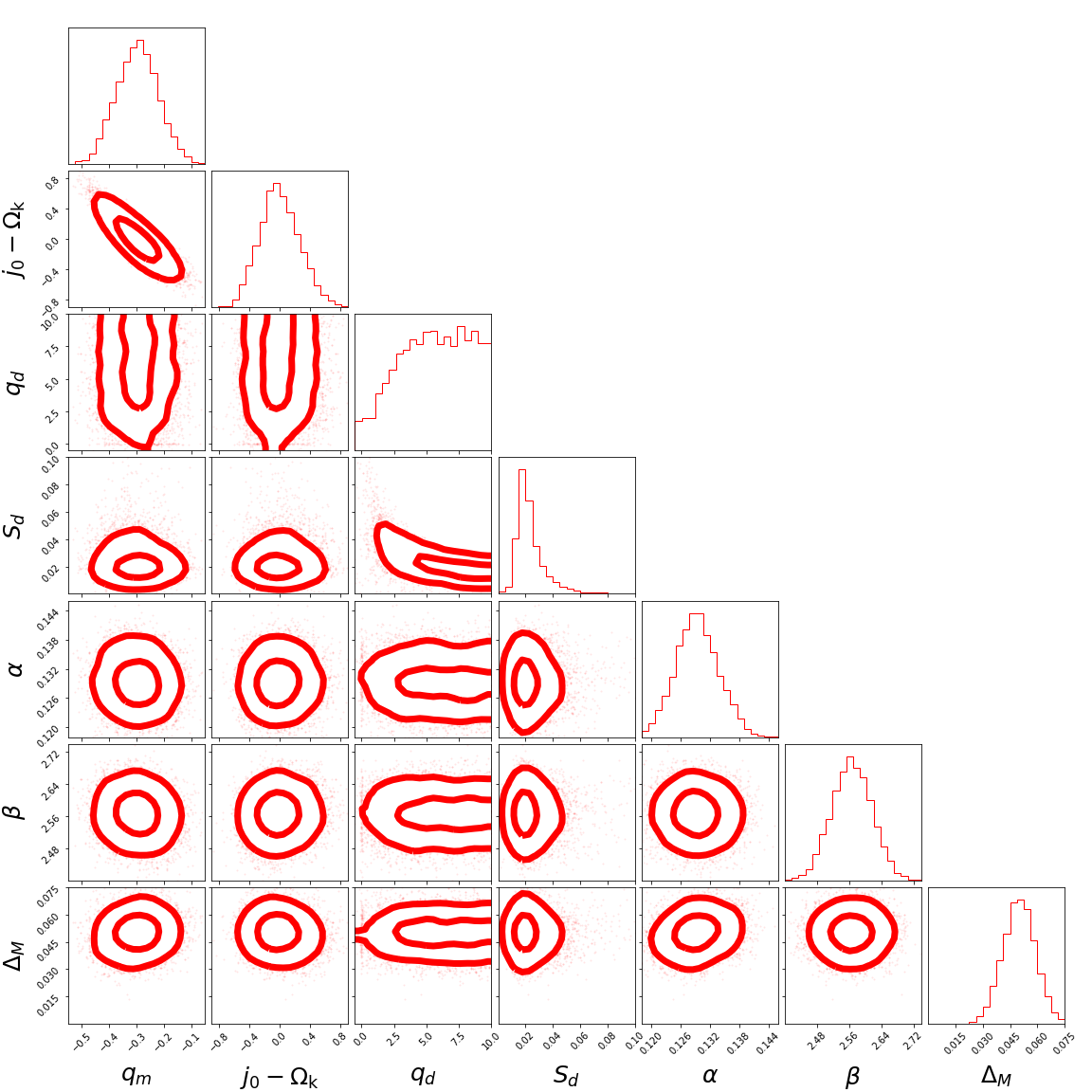}
    \caption{Posterior distribution of the cosmological and SN~Ia nuisance parameters using the Pantheon compilation. We find that cosmological parameters for the model are not correlated with the nuisance parameters for SN~Ia standardisation relations. The contours are 1- and 2-$\sigma$ respectively.}
    \label{fig:nuisance_params}
\end{figure*}
In Figure~\ref{fig:nuisance_params}, we present the posterior distribution for the cosmological parameters inferred from the Pantheon compilation using CMB frame redshifts, i.e. the monopole, dipole of the deceleration parameter, the decay scale for the dipole and the cosmological jerk minus curvature,  along with the nuisance parameters, $\alpha$, $\beta$, $\Delta_M$ and $M_B$. We find that the cosmological parameters, $q_m$, $q_d$ are uncorrelated with the SN~Ia standardisation parameters, using a Pearson $r$ test and finding $|r|$ values $< 0.1$. A similar correlation between the nuisance parameters and anisotropic cosmologies for the JLA compilation is presented in \citet{Dam:2017xqs, 2021arXiv210812497R}.

\section{Testing the direction of the dipole and observer boost}\label{appx:dip_direc} 

Our analysis presented in the main text is based on fixing the direction of the dipole in the deceleration parameter to that of the CMB dipole measured by \citet{2020A&A...641A...1P}. Further, in our {search for} %isotropic analysis to find 
the best-fit rest frame for us as observers --- i.e., not \textit{a priori} assuming this to be that of the CMB --- we also fix the direction of our velocity to coincide with that inferred from the CMB dipole. In this appendix, we present {a search for the optimal directions of these quantities across the sky.} %our tests of varying both of these directions across the sky, repeating the same fit, and determining the likelihood of each direction. 
For both tests,  we vary the direction of the dipole %vector ${\bf n}$ (of either the dipole in the deceleration parameter or the observer boost velocity) 
(associated with either the effective deceleration parameter or the observer boost velocity)
to coincide with indices of a \texttt{HEALPix} map with $N_{\rm side}=2$, i.e. $12\times N_{\rm side}^2 = 48$ directions in total. 

\begin{figure*}
    \centering
    \includegraphics[width=0.8\textwidth]{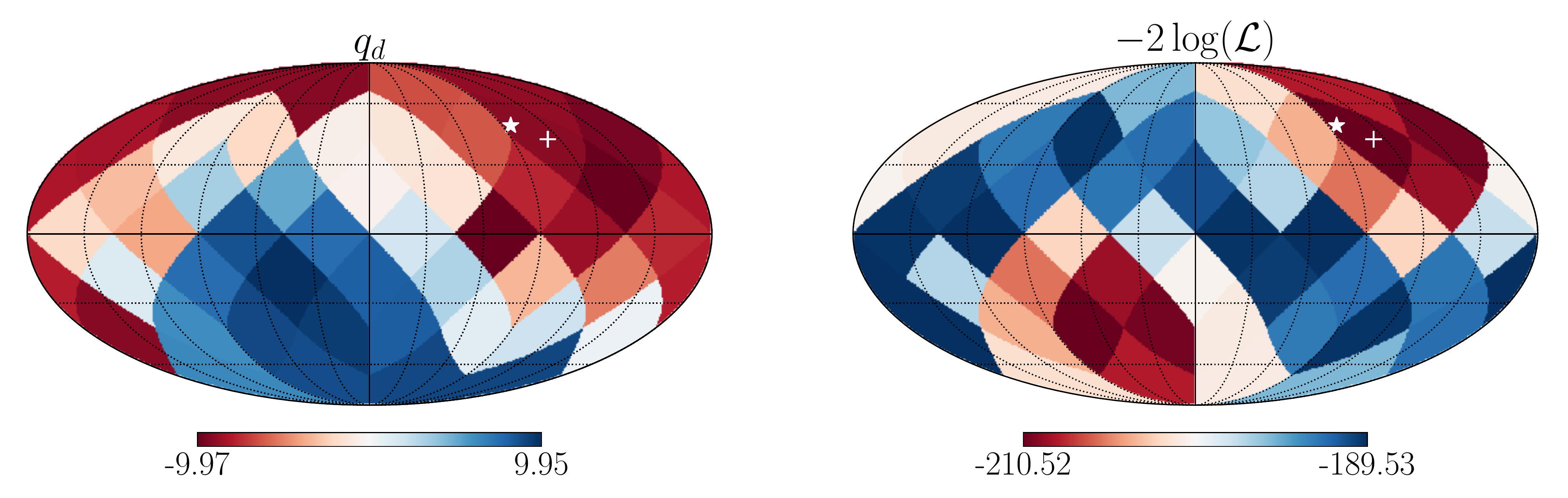}
    \caption{Left: norm of the deceleration dipole vector ${\bf q_d}$ as we vary the direction of the dipole to coincide with each \texttt{HEALPix} index shown here. %The negative values of $|{\bf q_d}|$ indicates a direction vector pointing in the opposite direction to the positive values. 
    Right: the likelihood of each fit performed here. We show $-2{\rm log}(\mathcal{L})$ for each instance of dipole direction vector. The white star in each panel represents the direction of the CMB dipole, and the white cross is the best-fit direction resulting from this test. }
    \label{fig:qd_direc}
\end{figure*}
\begin{figure*}
    \centering
    \includegraphics[width=0.8\textwidth]{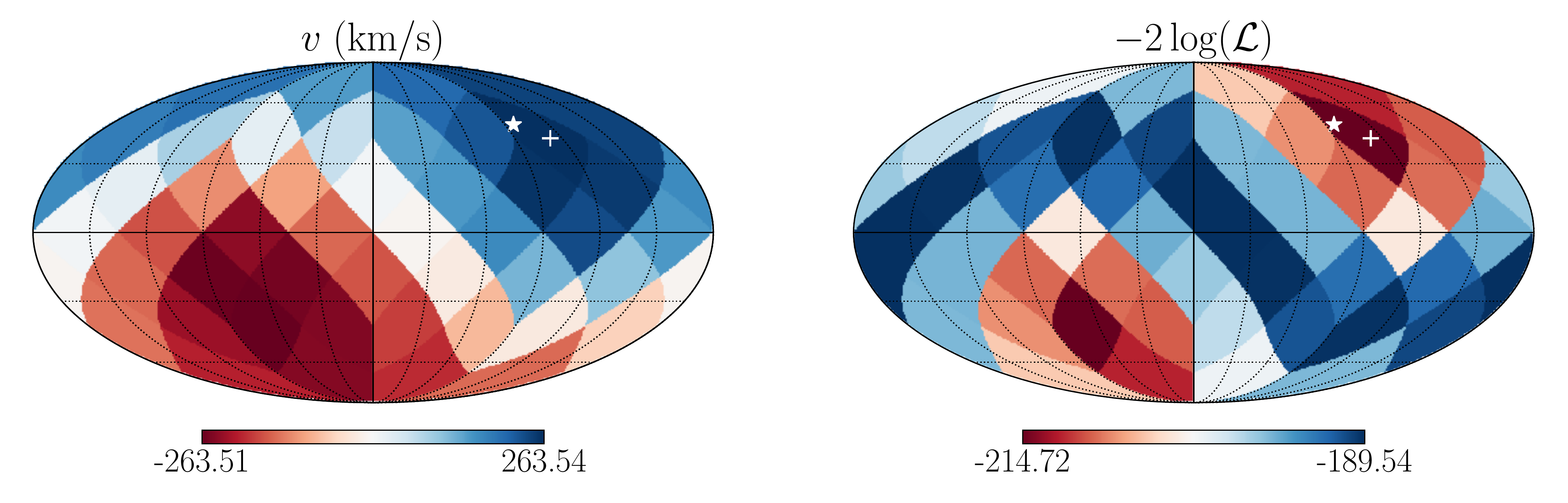}
    \caption{Left: norm of the velocity vector ${\bf v}$ as we vary the direction of the observer boost to coincide with each \texttt{HEALPix} index shown here. %The negative values of $|{\bf v}|$ indicates a direction vector pointing in the opposite direction to the positive values. 
    Right: the likelihood of each fit performed here. We show $-2{\rm log}(\mathcal{L})$ for each instance of velocity direction. The white star in each panel represents the direction of the CMB dipole, and the white cross is the best-fit direction resulting from this test. }
    \label{fig:v_direc}
\end{figure*} 

Figure~\ref{fig:qd_direc} shows the result of our test of the best-fit direction of the dipole in the deceleration parameter, namely, the {direction vector associated with}  %component 
{${\bf q_d} \equiv q_d {\bf n}$} in \eqref{eq:q_z_dipole}. 
The left panel shows a Mollweide projection of the {best fit} amplitude, {$q_d$}, %${\bf q_d}\cdot{\bf n}$, 
for each corresponding dipole direction, ${\bf n}$. %The directions with negative values of the norm indicate a direction opposite to those with positive values. 
The right panel of Figure~\ref{fig:qd_direc} shows the corresponding {profile log-}likelihood function{, $-2{\rm log}(\mathcal{L})$, for the direction}.  %values for each direction we have tested. 
%We show $-2{\rm log}(\mathcal{L})$, and so the minimum values on this scale correspond to the maximum likelihood direction. 
The white star on both panels corresponds to the direction of the CMB dipole from \citet{2020A&A...641A...1P}, and the white cross is {the} best-fit direction {of our analysis, corresponding to the minimum value of $-2{\rm log}(\mathcal{L})$}. % (which can be seen to coincide with the minimum value of $-2{\rm log}(\mathcal{L})$).

Figure~\ref{fig:v_direc} shows the result of our test of the best-fit direction of the observer boost. Specifically, we perform an isotropic cosmological fit with the velocity of the observer{, ${\bf v} = v {\bf n}$,} left as a {free variable}. 
{When redshifts are transformed to the CMB frame, ${\bf v}$ is chosen such that the dipole in the CMB temperature field vanishes, while here we leave the best fit rest frame to be determined from the SN~Ia catalogue itself.} 
The left panel shows a Mollweide projection of the {best fit} amplitude of the velocity, {$v$}, %${\bf v}\cdot{\bf n}$, 
for each corresponding direction ${\bf n}$. 
%\ahr{Commented out the sentence about negative values here. } 
%The directions with negative values of the norm indicate a direction opposite to those with positive values. 
The right panel of Figure~\ref{fig:v_direc} shows the corresponding {profile log-likelihood function for the direction}.  % likelihood values for each direction we have tested. 
The white star on both panels again corresponds to
the direction of the CMB dipole from \citet{2020A&A...641A...1P}, and the white cross is {the} best-fit direction {of our analysis}. % (which can be seen to coincide with the minimum value of $-2{\rm log}(\mathcal{L})$).

The best-fit direction {agrees well} %is identical 
between the fits for ${\bf q_d}$ and ${\bf v}${, with any difference within the angular resolution of our analysis}. 
%From this test, for both the dipole in the deceleration parameter and the boost velocity of the observer, we find the 
{This shared} best fit direction (white cross on all panels) closely coincides with the direction of the CMB dipole (white star on all panels).  \citet{Colin:2019} performed this same test for their fits for the dipole in the deceleration parameter, and found their best-fit direction to be $23$\textdegree away from the CMB dipole. %Due to our lower resolution ($N_{\rm side}=2$ as opposed to their $N_{\rm side}=8$), we cannot put a value on the distance between our best-fit direction and the CMB dipole direction because they fall within the same \texttt{HEALPix} cell. 
{We find our results to be} 
%However, this result is 
consistent with that of \citet{Colin:2019} given our resolution. 

\bsp	% typesetting comment
\label{lastpage}
\end{document}